\documentclass[]{article}
\usepackage{amsmath}
\usepackage{amsfonts}
\usepackage{amssymb}
\usepackage{siunitx}
\usepackage{graphicx}
\usepackage{physics}
\usepackage[affil-it]{authblk}
\usepackage{appendix}

\title{Super Heat Kernel of General Second Order Operators in $N=1$ Superspace and One-Loop Divergence of Dilaton-coupled SYM Theory\thanks{Work supported in part by the Director, Office of Science, Office of High Energy and Nuclear Physics, Division of High Energy Physics, of the US Department of Energy under Contract DE-AC02-05CH11231 and in part by the National Science Foundation under grant PHY-1316783.}}
\author{Ka-Hei Leung\thanks{Email: kaheileung@berkeley.edu}}
\affil{Department of Physics and Theoretical Physics Group,\\
	Lawrence Berkeley National Laboratory, University of California, \\
	Berkeley, California 94720, USA}
\date{\vspace{-4ex}}

\begin{document}

\maketitle

\begin{abstract}
We shall develop a general technique to obtain the super heat kernel coefficients of an arbitrary second order operator in $N=1$ superspace. We focus on the space of conformal supergravity here but the method presented is equally applicable for other types of superspace. The first three coefficients which determine the one-loop divergence of the corresponding quantum theory will be calculated. As an application we shall present the one-loop logarithmic divergence of super Yang-Mills theory coupled to a string dilaton $S$. This is the first superfield calculation for SYM with a non-trivial gauge kinetic function, which generalize the previous result with a constant coupling strength. We also demonstrate that the method presented can be extended to the case of third order operators, with the restriction that its third order part is composed of only spinor derivatives.
\end{abstract}

\newpage
\section{Introduction}\label{intro}
In the previous work \cite{Leung19}, we considered the super Yang-Mills theory in conformal supergravity and analyzed its one-loop effective action via the heat kernel method. We have developed a non-iterative technique which allows one to calculate the heat kernel coefficients efficiently. However, the previously considered model is restrictive, in the sense that the Yang-Mills coupling is a constant, in other words, the \emph{gauge kinetic function} $f_{(r)(s)}$ is trivial. A Yang-Mills theory with a non-trivial gauge kinetic function is of phenomenological interest, as in various supersymmetric models the guage coupling will be determined by the VEV of some, possibly composite, field. For instance, a four dimensionalal effective theory obtained from dimension reduction of a superstring theory will process a dynamical Yang-Mills gauge coupling \cite{Witten85}:
\begin{equation}\label{10dreducedYM}
\frac{1}{g^2}=e^{3\sigma}\phi^{-3/4}.
\end{equation}
Here $\phi$ is the string dilaton and $\sigma$ is a scalar field which emerges from the dimension reduction of the graviton.

In the following, we shall consider a simple case in which the gauge kinetic function is diagonal in the gauge index and is determined by a single dilaton field $S$. This typically arises from string theory models, for instance it may come from a weakly coupled heterotic string theory with orbifold compactification.\footnote{Studies on this type of theories can be found in \cite{Gaillard07} for example.} It will be seen that the previously presented non-recursive method is insufficient to calculate the heat kernel coefficients of the above scenario with a dilaton introduced. In fact, the issue is that the imposed constraints for the non-recursive method cannot be satisfied. To overcome the difficulties, we are going to develop an alternative technique, similar to the one in \cite{McArthur84}, to calculate the heat kernel coefficients in this case. The method presented here actually applies to any second order operators, thus potentially has a broad class of applications. 

In this work, we will start with a discussion of the super Yang-Mills theory with a string dilaton. We are going to obtain the operator that determines the one-loop effective action. We shall work in the conformal superspace developed by Butter \cite{Butter10}, but by suitably fixing the conformal symmetry we also obtain the case in $U(1)$-supergravity as discussed in \cite{BBG00} or in the more familiar minimal supergravity. Next we will develop a technique that enables the calculation of heat kernel coefficients of an arbitrary second operator $\mathcal{O}$. The first three coefficients will be presented here. Then we will apply the general result to the case of SYM with a dilaton, and derive its one-loop logarithmic divergence. In the final section, we shall briefly argue that the method here applies to a certain class of third order operators, in which the third order part contains only spinor derivatives.

\section{Super Yang-Mills with a Dilaton}\label{DilSYM1}

In this section, we shall consider super Yang-Mills theory in $N=1$ conformal supergravity, with the gauge kinetic function determined by a dilaton field $S$. We will work with the superfield approach of conformal supergravity, developed in \cite{Butter10} and briefly reviewed in \cite{Leung19}. We will quantize this theory and eventually calculate the operator that encodes the one-loop effective action, and thus its divergence, of the vector multiplet. The treatment here will be similar to the constant coupling case, which was previously considered \cite{Leung19}. A review of the conformal superspace and the quantization of SYM with constant coupling can be found in the appendix.

\subsection{Quantization of the Theory}\label{DilSYM1.1}
Let us start with the classical action
\begin{equation}\label{DilYMaction}
S_{\textrm{YM},S}=\frac{1}{4}\int d^4xd^2\theta \, \mathcal{E} \tr\left(S\mathcal{W}_{\textrm{YM}}^{\alpha}\mathcal{W}^{}_{\textrm{YM}\alpha}\right)+\textrm{h.c.},
\end{equation}
where $S$ is the string dilaton field, which corresponds to a gauge kinetic function $f_{(r)(s)}=S\delta_{(r)(s)}$. Here $S$ is a chiral primary field with vanishing conformal weight, and it is a Yang-Mills gauge singlet. Obviously a constant coupling is just the special case: $S\rightarrow1/g^2$. In general, $S$ has a non-trivial spacetime dependence, and is complex. The treatment for an even more general setup will be similar.

It is easy to see that one can define the vector multiplet, or a scalar superfield, $V$ exactly as the constant coupling case \cite{Leung19}, which will give us the second order action with only minor modifications needed:
\begin{equation}\label{DilYM2nd}
S^{(2)}_{\textrm{YM},S}=\frac{1}{16}\int d^4xd^4\theta \, ES\tr\left(\nabla^{\alpha}V\bar{\nabla}^{2}\nabla_{\alpha}V-4\mathcal{W}_{\textrm{YM}}^\alpha[V,\nabla_{\alpha}V]\right)+\textrm{h.c.}.
\end{equation}
As for gauge fixing, we shall have the same gauge-fixing functional as before: $f=\bar\nabla^2(XV)$, and its conjugate. The gauge fixing action can be found by substituting $2/g^2\rightarrow S+\bar{S}$ in the one used in \cite{Leung19}:
\begin{equation}\label{Dilgfaction}
S^{(V)}_{\textrm{g.f.}}=\frac{1}{16}\tr\int d^8z \,E(S+\bar{S})X^{-2}[\bar\nabla^2(XV)\nabla^2(XV)].
\end{equation}
Since we have the identical gauge-fixing functional as the old case, the Faddeev-Popov ghosts receive no change:
\begin{equation}\label{DilFPaction}
S_{\textrm{FP}}=\tr\int d^8z \, EX(c'+\bar{c}')\mathcal{L}_{V/2}[c-\bar{c}+\coth(\mathcal{L}_{V/2})(c+\bar{c})].
\end{equation}
However the Nielsen-Kallosh ghost will develop a dilaton dependence:
\begin{equation}\label{DilNKaction}
S_{\textrm{NK}}=\tr\int d^8z \, EX^{-2}(S+\bar{S})b\bar{b},
\end{equation}
which is seen from the appearance of $S$ in \eqref{Dilgfaction}.

\subsection{Second Order Action}\label{DilSYM1.2}
We have to simplify the second order vector multiplet action into the form \begin{equation}\label{simplifyDilaction2nd}
S_{\textrm{YM},S}^{(2)}=\frac{1}{2}\int d^8z\, E\tr(V\mathcal{O}_{V,S}V),
\end{equation}
which determines the one-loop effective action. The procedure is similar to the trivial kinetic function case, except that derivatives of the dilaton field will appear. For example, for the term $S\nabla^{\alpha}V\bar{\nabla}^{2}\nabla_{\alpha}V$, using integration by parts, which is non-trivial as discussed in the appendix, we have 
\begin{equation}\label{DILYMbyparts1}
S\nabla^{\alpha}V \bar{\nabla}^{2}\nabla_{\alpha}V=-SV\nabla^{\alpha} \bar{\nabla}^{2}\nabla_{\alpha}V-V\nabla^\alpha S\bar{\nabla}^{2}\nabla_{\alpha}V,
\end{equation}
here $SV \bar{\nabla}^{2}\nabla_{\alpha}V$ is primary so there is no integration by parts correction. The expression with $\nabla^\alpha S$ is a new term which only appears when a dilaton is introduced, which has three derivatives acting on $V$. In the constant coupling case, the first term with four derivatives will be canceled by the gauge-fixing term, so only terms with less than two derivatives survive. In the dilaton setup, we will see that the four derivative terms will be again canceled, but the extra new term will remain. Thus we potentially have to deal with a differential operator containing terms with three derivatives, which requires careful analysis.

Next we have the term $-4S\mathcal{W}_{\textrm{YM}}^\alpha[V,\nabla_{\alpha}V]$, and some algebra gives: 
\begin{equation}\label{DilYMbyparts2}
\begin{aligned}
&-4S\tr\mathcal{W}_{\textrm{YM}}^\alpha[V,\nabla_{\alpha}V]\\ 
=&\,4\tr\nabla^{\alpha}VS\mathcal{W}_{\textrm{YM},\alpha}V+4\tr SV\mathcal{W}_{\textrm{YM}}^\alpha\nabla_{\alpha}V\\
=&\,-4\tr V\nabla^\alpha(S\mathcal{W}_{\textrm{YM},\alpha}V)+4\tr SV\mathcal{W}_{\textrm{YM}}^\alpha\nabla_{\alpha}V\\
=&\,4\tr V[2S\mathcal{W}_{\textrm{YM}}^\alpha\nabla_{\alpha}-S(\nabla^\alpha\mathcal{W}_{\textrm{YM},\alpha})-\nabla^\alpha S\mathcal{W}_{\textrm{YM},\alpha}]V.
\end{aligned}
\end{equation}
To go from the first line to the second, cyclicity of traces is used. Then integration by parts is applied on the first term to get the third line. Note that there is no correction term as every object appearing is primary.

We now turn our attention to the gauge-fixing term. Similar to the trivial kinetic function case, we use cyclicity of traces to symmetrically split the term into two:
\begin{equation}\label{DilgfVaction}
S^{(V)}_{\textrm{g.f.}}=\frac{1}{16}\tr\int d^8z \,E\chi X^{-2}[\bar\nabla^2(XV)\nabla^2(XV)+\nabla^2(XV)\bar\nabla^2(XV)]
\end{equation}
we have defined $\chi=\frac{1}{2}(S+\bar{S})$, the real part of $S$, for convenience. Let us recall some of the adopted notation in \cite{Leung19}, which will be used here also:
\begin{equation}\label{DilYMdefrecall1}
\begin{aligned}
Y=X^{-1}\bar\nabla^2(XV)&=(\bar\nabla^2+2U_{\dot{\alpha}}\nabla^{\dot{\alpha}}-8R)V\\
\bar{Y}=X^{-1}\nabla^2(XV)&=(\nabla^2+2U^\alpha\nabla_{\alpha}-8\bar{R})V\\
U^\alpha=\nabla^{\alpha}\log{X}, &\quad U_{\dot{\alpha}}=\nabla_{\dot\alpha}\log{X},\\
R=-\frac{1}{8X}\bar\nabla^2X, &\quad\bar{R}=-\frac{1}{8X}\nabla^2X,\\
X_\alpha=\frac{3}{8}\bar\nabla^2U_{\alpha}, &\quad X^{\dot{\alpha}}=\frac{3}{8}\nabla^2U^{\dot\alpha},\\
G_{\alpha\dot{\alpha}}=-\frac{1}{4}(U_{\alpha\dot{\alpha}}&-U_{\dot{\alpha}\alpha})-\frac{1}{2}U_{\alpha}U_{\dot{\alpha}},\\
U_{\alpha\dot{\alpha}}=\nabla_{\alpha}U_{\dot{\alpha}}, &\quad U_{\dot{\alpha}\alpha}=\nabla_{\dot\alpha}U_{\alpha}, \quad \textrm{etc.}
\end{aligned}
\end{equation}

We shall employ integration by parts on the term $\bar\nabla^2V(\chi\bar{Y})$, and it is not hard to see that 
\begin{equation}\label{DilYMbyparts3}
\begin{aligned}
&\bar\nabla^2V(\chi\bar{Y})\\
=\,&-\nabla_{\dot{\alpha}}V\nabla^{\dot{\alpha}}(\chi\bar{Y})\\
=\,&8f_{\dot{\alpha}}{}^{\dot{\alpha}}V\chi\bar{Y}+V\bar\nabla^2(\chi\bar{Y}).
\end{aligned}
\end{equation}
There is an integration by parts correction containing the special conformal connection $f_A{}^B$, which can be found the same way as for the constant coupling case. Expanding the term $\bar\nabla^2(\chi\bar{Y})$ gives us
\begin{equation}\label{DilYMbyparts3a}
\begin{aligned}
&\bar\nabla^2V(\chi\bar{Y})\\
=\,&8f_{\dot{\alpha}}{}^{\dot{\alpha}}V\chi\bar{Y}+\chi V\bar\nabla^2\bar{Y}+ 2V\nabla_{\dot\alpha}\chi\nabla^{\dot{\alpha}}\bar{Y}+ V\bar\nabla^2\chi\bar{Y}\\
=\,&8f_{\dot{\alpha}}{}^{\dot{\alpha}}V\chi\bar{Y}+\chi V\bar\nabla^2\bar{Y}+ V\nabla_{\dot\alpha}\bar{S}\nabla^{\dot{\alpha}}\bar{Y}+ V\bar\nabla^2\bar{S}\bar{Y}/2,
\end{aligned}
\end{equation}
where we have used the chirality of $S$: $\nabla_{\dot{\alpha}}\chi=\nabla_{\dot{\alpha}}\bar{S}/2$.

The next term we consider is $2U_{\dot{\alpha}}\nabla^{\dot{\alpha}}V(\chi\bar{Y})$. Integration by parts gives
\begin{equation}\label{DilYMbyparts4}
\begin{aligned}
&2U_{\dot{\alpha}}\nabla^{\dot{\alpha}}V(\chi\bar{Y})\\
=\,&-8f_{\dot{\alpha}}{}^{\dot{\alpha}}V\chi\bar{Y}-2V\nabla_{\dot\alpha}(U^{\dot{\alpha}}\chi\bar{Y})\\
=\,&-8f_{\dot{\alpha}}{}^{\dot{\alpha}}V\chi\bar{Y}-2\chi V\nabla_{\dot\alpha}(U^{\dot{\alpha}}\bar{Y})-V\nabla_{\dot\alpha}\bar{S} U^{\dot{\alpha}}\bar{Y}.
\end{aligned}
\end{equation}
Note that the two correction terms cancel as in the case without a dilaton. Combine with $-8R\chi\bar{Y}$, and notice that the terms without derivatives of $\chi$ were previously encountered in the constant coupling case. After some work we get the expression
\begin{equation}\label{DilYMbyparts5}
\begin{aligned}
&\chi Y\bar{Y}\\
=\,&V\chi\bar\nabla^2\nabla^2V+V[2\chi U^\alpha\bar\nabla^2\nabla_{\alpha}+(\nabla_{\dot\alpha}\bar{S}-2\chi U_{\dot{\alpha}})\nabla^{\dot{\alpha}}\nabla^2]V\\
&+V\left[(8\chi R+2\chi U_{\dot{\alpha}}U^{\dot{\alpha}}+\bar{\nabla}^2\bar{S}/2)\nabla^2-8\chi\bar{R}\bar\nabla^2\right.\\
&+\left.(4\chi U^{\dot{\alpha}\alpha}-4\chi U^{\dot{\alpha}}U^{\alpha}
+\nabla^{\dot{\alpha}}\bar{S}U^\alpha)\nabla_{\dot\alpha}\nabla_{\alpha}\right]V\\
&+V\left(\frac{16}{3}\chi X^{\alpha}+16\chi RU^\alpha+4\chi U_{\dot{\alpha}}U^{\dot{\alpha}}U^\alpha+\nabla_{\dot\alpha}\bar{S}U^{\dot{\alpha}\alpha}+\bar\nabla^2\bar{S}U^\alpha\right.\\
&\left.-4\chi U_{\dot\alpha}U^{\dot{\alpha}\alpha}-2\nabla_{\dot\alpha}\bar{S}U^{\dot{\alpha}}U^{\alpha}\right)\nabla_{\alpha}V+8V(2\chi\bar{R}U_{\dot{\alpha}}-2\chi\nabla_{\dot\alpha}\bar{R}-\nabla_{\dot\alpha}\bar{S}\bar{R})\nabla^{\dot{\alpha}}V\\
&-8V(\chi\bar\nabla^2\bar{R}+8\chi R\bar{R}+2\chi U_{\dot{\alpha}}U^{\dot{\alpha}}\bar{R}+\nabla_{\dot{\alpha}}\bar{S}\nabla^{\dot{\alpha}}\bar{R}+\bar\nabla^2\bar{S}\bar{R}/2\\
&-2\chi U_{\dot{\alpha}}\nabla^{\dot{\alpha}}\bar{R}-\nabla_{\dot{\alpha}}\bar{S}U^{\dot{\alpha}}\bar{R})V.
\end{aligned}
\end{equation}
We also have
\begingroup
\allowdisplaybreaks
\begin{align*}
&\chi Y\bar{Y}+\chi\bar{Y}Y\\
=\,&V\chi(\bar\nabla^2\nabla^2+\nabla^2\bar\nabla^2)V+2V\chi(U^\alpha\left[\bar\nabla^2,\nabla_{\alpha}\right]+U_{\dot{\alpha}}\left[\nabla^2,\nabla^{\dot{\alpha}}\right])V\\
&+V(\nabla_{\dot\alpha}\bar{S}\nabla^{\dot{\alpha}}\nabla^2+\nabla^{\alpha}S\nabla_{\alpha}\bar\nabla^2)V\\
&+V[(\bar\nabla^2\bar{S}/2+2\chi U_{\dot{\alpha}}U^{\dot{\alpha}})\nabla^2+(\nabla^2S/2+2\chi U^{\alpha}U_{\alpha})\bar\nabla^2]V\\
&+V(8\chi G^{\alpha\dot{\alpha}}+8\chi U^{\alpha}U^{\dot{\alpha}}+\nabla^{\dot{\alpha}}\bar{S}U^\alpha/2-\nabla^\alpha SU^{\dot{\alpha}}/2)[\nabla_{\dot\alpha,}\nabla_{\alpha}]V\\
&+V\left(\frac{16}{3}\chi X^{\alpha}-16\chi\nabla^\alpha{R}-8\nabla^{\alpha}SR+32\chi RU^\alpha+4\chi U_{\dot{\alpha}}U^{\dot{\alpha}}U^\alpha\right.\\
&+\left.\nabla_{\dot\alpha}\bar{S}U^{\dot{\alpha}\alpha}+\bar\nabla^2\bar{S}U^\alpha-4\chi U_{\dot\alpha}U^{\dot{\alpha}\alpha}-2\nabla_{\dot\alpha}\bar{S}U^{\dot{\alpha}}U^{\alpha}\right)\nabla_{\alpha}V\\ 
&+V\left(\frac{16}{3}\chi X_{\dot{\alpha}}-16\chi\nabla_{\dot\alpha}\bar{R}-8\nabla_{\dot\alpha}\bar{S}\bar{R}+32\chi\bar{R}U_{\dot{\alpha}}+4\chi U^{\alpha}U_{\alpha}U_{\dot{\alpha}}\right.\\
&\left.+\nabla^{\alpha}SU_{\alpha\dot{\alpha}}+\nabla^2SU_{\dot{\alpha}}-4\chi U^{\alpha}U_{\alpha\dot{\alpha}}-2\nabla^\alpha SU_\alpha U^{\dot{\alpha}}\right)\nabla^{\dot{\alpha}}V\\
&+[16VU^a-i\sigma^a_{\alpha\dot{\alpha}}(\nabla^{\dot{\alpha}}\bar{S}U^\alpha+\nabla^\alpha SU^{\dot{\alpha}})]\nabla_aV-8V(\chi\bar\nabla^2\bar{R}+\chi\nabla^2R\\
&+16\chi R\bar{R}+2\chi U_{\dot{\alpha}}U^{\dot{\alpha}}\bar{R}+2\chi U^{\alpha}U_{\alpha}R+\nabla_{\dot{\alpha}}\bar{S}\nabla^{\dot{\alpha}}\bar{R}+\nabla^\alpha S\nabla_{\alpha}R+\bar\nabla^2\bar{S}\bar{R}/2\\
&+\nabla^2SR/2-2\chi U^{\alpha}\nabla_{\alpha}R-2\chi U_{\dot{\alpha}}\nabla^{\dot{\alpha}}\bar{R}-\nabla_{\dot{\alpha}}\bar{S}U^{\dot{\alpha}}\bar{R}-\nabla^\alpha SU_\alpha R)V.
\addtocounter{equation}{1}\tag{\theequation}\label{DilYMbyparts6}
\end{align*}
\endgroup

We would like to remove the terms with too many derivatives using the following identities, listed in \cite{Kugo16a}:
\begin{equation}\label{DilYMkeyidentity1}
\begin{aligned}
\nabla^2\bar\nabla^2+\bar\nabla^2\nabla^2-\nabla^\alpha\bar\nabla^2\nabla_{\alpha}-\nabla_{\dot\alpha}\nabla^2\nabla^{\dot{\alpha}}&=16\Box+8\mathcal{W}^\alpha\nabla_{\alpha}-8\mathcal{W}_{\dot{\alpha}}\nabla^{\dot{\alpha}},\\
\nabla^\alpha\bar\nabla^2\nabla_{\alpha}-\nabla_{\dot\alpha}\nabla^2\nabla^{\dot{\alpha}}&=8(\mathcal{W}^\alpha\nabla_{\alpha}+\mathcal{W}_{\dot{\alpha}}\nabla^{\dot{\alpha}}+\{\nabla^\alpha,\mathcal{W}_{\alpha}\}),\\
\left[\bar\nabla^2,\nabla_{\alpha}\right]&=2i\nabla^{\dot{\beta}}\nabla_{\alpha\dot{\beta}}+2i\nabla_{\alpha\dot{\beta}}\nabla^{\dot{\beta}},\\
\left[\nabla^2,\nabla^{\dot\alpha}\right]&=2i\nabla^{\alpha\dot{\beta}}\nabla_{\alpha}+2i\nabla_{\alpha}\nabla^{\alpha\dot{\beta}}.
\end{aligned}
\end{equation}
The first two equations imply that 
\begin{equation}\label{DilYMkeyidentity2}
\begin{aligned}
&\chi\nabla^2\bar\nabla^2+\chi\bar\nabla^2\nabla^2-S\nabla^\alpha\bar\nabla^2\nabla_{\alpha}-\bar{S}\nabla_{\dot\alpha}\nabla^2\nabla^{\dot{\alpha}}\\
=&16\chi\Box+8\bar{S}\mathcal{W}_{\textrm{YM}}^\alpha\nabla_{\alpha}-8S\mathcal{W}_{\textrm{YM},\dot{\alpha}}\nabla^{\dot{\alpha}}+4(S-\bar{S})\nabla^{\alpha}\mathcal{W}_{\textrm{YM},\alpha}.
\end{aligned}
\end{equation}
Note that we have replaced the gaugino $\mathcal{W}$ by its Yang-Mills part, as the other parts will vanish when acting on the vector multiplet $V$. We see that this equation allows the removal of terms with four derivatives, and using the Bianchi identity $\nabla^\alpha\mathcal{W}_{\textrm{YM},\alpha}=\nabla_{\dot\alpha}\mathcal{W}_{\textrm{YM}}^{\dot{\alpha}}$ the last term in \eqref{DilYMkeyidentity2} cancels with similar terms in \eqref{DilYMbyparts2} and its conjugate. For the terms with three derivatives, the third and the fourth equation in \eqref{DilYMkeyidentity1} can be used. The final result is that we have no terms with more than two derivatives, which is somewhat surprising as one might expect terms with three derivatives like $V\nabla^\alpha S\bar{\nabla}^{2}\nabla_{\alpha}V$ to persist, but the gauge-fixing term provides cancellation.

To conclude, we have, in the presence of a dilaton, the second order action
\begin{equation}\label{DilYMaction2nd}
S_{\textrm{YM},S}^{(2)}=\frac{1}{2}\tr\int d^8z\, EV\mathcal{O}_{V,S}V,
\end{equation}
with the operator $\mathcal{O}_{V,S}$, which determines the one-loop effective action, given by
\begin{equation}\label{DilYMOV}
\begin{aligned}
\mathcal{O}_{V,S}=&(S+\bar{S})\mathcal{O}_V+\frac{\bar\nabla^2\bar{S}}{16}\nabla^2+\frac{\nabla^2 S}{16}\bar{\nabla}^2\\
&-\frac{i}{4}\nabla^\alpha{S}(\nabla^{\dot{\beta}}\nabla_{\alpha\dot{\beta}}+\nabla_{\alpha\dot{\beta}}\nabla^{\dot{\beta}})-\frac{i}{4}\nabla_{\dot{\alpha}}\bar{S}(\nabla^{\beta\dot{\alpha}}\nabla_{\beta}+\nabla_{\beta}\nabla^{\beta\dot{\alpha}})\\
&+(\nabla_{\dot{\alpha}}\bar{S}U^{\dot{\alpha}\alpha}-\nabla^{\alpha}SR)\nabla^\alpha+(\nabla^\alpha SU_{\alpha\dot{\alpha}}-\nabla_{\dot\alpha}\bar{S}\bar{R})\nabla^{\dot\alpha}\\
&-(\nabla_{\dot{\alpha}}\bar{S}\nabla^{\dot{\alpha}}\bar{R}+\nabla^\alpha S\nabla_{\alpha}R+\bar\nabla^2\bar{S}\bar{R}/2+\nabla^2SR/2\\
&+\nabla^\alpha S\mathcal{W}_{\textrm{YM},\alpha}/2-\nabla_{\dot{\alpha}}\bar{S}\mathcal{W}^{\dot{\alpha}}_{\textrm{YM}}/2)\\
&+\frac{1}{16}(\nabla^{\dot{\alpha}}\bar{S}U^\alpha-\nabla^\alpha SU^{\dot{\alpha}})[\nabla_{\dot\alpha,}\nabla_{\alpha}]-\frac{i}{8}\sigma^a_{\alpha\dot{\alpha}}(\nabla^{\dot{\alpha}}\bar{S}U^\alpha+\nabla^\alpha SU^{\dot{\alpha}})\nabla_a\\
&+\frac{1}{8}(\bar\nabla^2\bar{S}U^{\alpha}-2\nabla_{\dot\alpha}\bar{S}U^{\dot\alpha}U^{\alpha})\nabla_{\alpha}+\frac{1}{8}(\nabla^2SU_{\dot{\alpha}}-2\nabla^\alpha SU_\alpha U^{\dot{\alpha}})\nabla^{\dot{\alpha}}\\
&+\nabla_{\dot{\alpha}}\bar{S}U^{\dot{\alpha}}\bar{R}+\nabla^\alpha SU_\alpha R
\end{aligned}
\end{equation}
where $\mathcal{O}_V$ is the operator that corresponds to the case of a trivial gauge kinetic function, which was derived previously \cite{Leung19}. We recall it here for completeness:
\begin{equation}\label{YMOVrecall}
\begin{aligned}
\mathcal{O}_V&=\Box+\frac{1}{2}G^{\alpha\dot{\alpha}}[\nabla_{\alpha},\nabla_{\dot\alpha}]+\left(\frac{X^\alpha}{3}-\nabla^{\alpha}R+\mathcal{W}_{\textrm{YM}}^{\alpha}\right)\nabla_{\alpha}\\
&+\left(\frac{X_{\dot\alpha}}{3}-\nabla_{\dot\alpha}\bar{R}-\mathcal{W}_{\textrm{YM},\dot\alpha}\right)\nabla^{\dot{\alpha}}-\frac{1}{2}\left(\bar\nabla^2\bar{R}+\nabla^2R+16R\bar{R}\right)\\
&+\frac{i}{4}U^\alpha(\nabla^{\dot{\beta}}\nabla_{\alpha\dot{\beta}}+\nabla_{\alpha\dot{\beta}}\nabla^{\dot{\beta}})+\frac{i}{4}U_{\dot{\alpha}}(\nabla^{\beta\dot{\alpha}}\nabla_{\beta}+\nabla_{\beta}\nabla^{\beta\dot{\alpha}})\\
&+\frac{1}{8}\left(U_{\dot{\alpha}}U^{\dot{\alpha}}\nabla^2+U^{\alpha}U_{\alpha}\bar\nabla^2+4U^{\alpha}U^{\dot{\alpha}}[\nabla_{\alpha},\nabla_{\dot\alpha}]\right)\\
&+\frac{1}{4}\left(8RU^\alpha+U_{\dot{\alpha}}U^{\dot{\alpha}}U^\alpha-U_{\dot\alpha}U^{\dot{\alpha}\alpha}\right)\nabla_\alpha\\
&+\frac{1}{4}\left(8\bar{R}U_{\dot{\alpha}}+U^{\alpha}U_{\alpha}U_{\dot{\alpha}}-U^{\alpha}U_{\alpha\dot{\alpha}}\right)\nabla^{\dot{\alpha}}\\
&+U^a\nabla_a+\left(U^{\alpha}\nabla_{\alpha}R+U_{\dot{\alpha}}\nabla^{\dot{\alpha}}\bar{R}-U^{\alpha}U_{\alpha}R-U_{\dot{\alpha}}U^{\dot{\alpha}}\bar{R}\right).
\end{aligned}
\end{equation}
We also split the part that depends on derivatives of $S$ or $\bar{S}$ into two, one part that is not vanishing when setting the conformal gauge $U_A=0$, and one that vanishes.

It is clear that when $S=\bar{S}=g^{-2}$, we return to the old case as derivatives of the dilaton vanish. Note that by direct inspection, the leading term of $\mathcal{O}_{V,S}$ is $\mathcal{O}_{V,S}=(S+\bar{S})\Box+\cdots$. We still have a d'Alembertian as expected, but the coefficient $S+\bar{S}$ implies that the spacetime propagation of $V$ is influenced by the presence of the dilaton, which will need extra consideration.

\section{Heat Kernel as a Fourier Integral}\label{DilSYM2}
We have determined how the introduction of a dilaton affects the operator $\mathcal{O}_V$ governing the one-loop effective action. The next goal will be calculating how this changes the heat kernel coefficients. Previously in the case of constant coupling, we applied the de Witt heat kernel expansion and developed a non-recursive technique that allows us to calculate the heat kernel coefficients. It turns out that such a method is inadequate for the new scenario we have, one reason being that we have a non-trivial dependence in the d'Alembertian term: $\mathcal{O}_V=(S+\bar{S})\Box+\cdots$. Such $\mathcal{O}_V$ is classified as a \emph{non-minimal} operator, its treatment is more complicated than the minimal case, where the pre-factor of $\Box$ is absent. Analysis of heat kernel coefficients for non-minimal operators, especially for non-supersymmetric ones, has been studied using various methods, one example being \cite{Iochum17}. There were also one-loop studies of non-minimal operators in non-minimal supergravity \cite{Buchbinder89}, however an indirect method was employed and a direct calculation was not given. In the following, we shall develop a direct method, by employing a technique involving Fourier integrals that is applicable in a generic superspace, which was first demonstrated by McArthur \cite{McArthur84}.

\subsection{Expression for Heat Kernel Coefficients}\label{DilSYM2.1}
Recall that the super heat kernel $K$ of an operator $\mathcal{O}$ is defined by a differential equation and has the formal expression
\begin{equation}\label{recalldefshk}
\begin{aligned}
&\left(\mathcal{O}+i\frac{\partial}{\partial \tau}\right)K(z,z';\tau)=0,\\
&K(z,z';\tau)=e^{i\tau\mathcal{O}}E^{-1}\delta^8(z-z').
\end{aligned}
\end{equation}
It is possible to expand the heat kernel into a power series in $\tau$. In de Witt's approach, the expansion is of the form:
\begin{equation}\label{deWittHKexpand}
K(z,z';\tau)=\frac{-i}{(4\pi \tau)^2}\sum_{n=0}^\infty\exp\left(i\frac{\sigma}{2\tau}\right)\Delta^{1/2} a_n\frac{(i\tau)^n}{n!}.
\end{equation}
A superspace version of de Witt's method was developed in \cite{Buchbinder86}. However technical complications arise in the supersymmetric generalization. For example, it only works for a minimal operator, as in the non-supersymmetric case, thus it would be insufficient here. Another example is that one has to define an operator-dependent $\sigma$ and $\Delta$, as no natural metric exists in a superspace. We would like to avoid introducing operator-dependent objects, thus we shall turn to a different approach.

Here instead we will consider a slightly different expansion:
\begin{equation}\label{bnHKexpand}
K(z,z';\tau)=\frac{-i}{(4\pi \tau)^2}\sum_{n=0}^\infty b_n\frac{(i\tau)^n}{n!},
\end{equation}
without the object $\exp\left(i\frac{\sigma}{2\tau}\right)\Delta^{1/2}$. Note that for the one-loop effective action, we care about the \emph{coincidence limit} $[K(\tau)]=K|_{z'\rightarrow z}$, and the conditions $[\sigma]=0$ and $[\Delta]=1$ imply the two sets of coefficients share the same limit $[a_n]=[b_n]$.

For convenience, we set $z$ to be the superspace origin. As we will take the coincidence limit, it suffices to consider $z'$ to be near the origin, for which we may choose a normal coordinate system: $y^M=(y^m,y^\mu,y_{\dot{\mu}})$ \cite{McArthur84b}. Using such coordinates, it can be shown that \cite{McArthur84} the delta function appearing in \eqref{recalldefshk} has an integral representation:
\begin{equation}\label{Fouriersdelta}
\delta^8(z')=\int \frac{d^4k}{(2\pi)^4}\exp(iy^m\delta_{m}{}^{a}k_a)y^\mu y_{\mu}y_{\dot{\mu}}y^{\dot{\mu}}.
\end{equation}
This allows us to write, using the operator expression in \eqref{recalldefshk},
\begin{equation}\label{Fouriershk}
K(\tau)=\int \frac{d^4k}{(2\pi)^4}e^{i\tau\mathcal{O}}\exp(iy^m\delta_{m}{}^{a}k_a)E^{-1}y^\mu y_{\mu}y_{\dot{\mu}}y^{\dot{\mu}}.
\end{equation}

In the following, we would like to calculate the coincidence limit of such an integral, thus obtaining the coefficients $[b_n]$. We for now restrict ourselves to the case of $\mathcal{O}$ being a second order differential operator, with terms at most quadratic in covariant derivatives. This in particular covers the case of super Yang-Mills with a dilaton, which is our main interest. We will see that it is possible to generalize such a method to some special cases in which higher derivative terms may appear.

Let us define $\phi=iy^m\delta_{m}{}^{a}k_a$, and we want to move the factor $e^\phi$ in \eqref{Fouriershk} past the operator $e^{i\tau\mathcal{O}}$. This can be achieved by using the operator identity
\begin{equation}\label{econjugateidentity}
e^\lambda \chi e^{-\lambda}=e^{\mathcal{L}_\lambda}\chi,
\end{equation}
where $\mathcal{L}_\lambda\chi=[\lambda,\chi]$ is the commutator. This identity can be seen straightforwardly by Taylor expanding the exponentials and checking that both sides are equal order by order in $\lambda$. \eqref{econjugateidentity} implies
\begin{equation}\label{Fouriershk2}
K=\int \frac{d^4k}{(2\pi)^4}e^\phi\exp\left(\sum_{m=0}^{\infty}\frac{(-1)^m}{m!}(\mathcal{L}_\phi)^m(i\tau\mathcal{O})\right)E^{-1}y^\mu y_{\mu}y_{\dot{\mu}}y^{\dot{\mu}}.
\end{equation}
For a second order operator $\mathcal{O}$, $(\mathcal{L}_\phi)^m\mathcal{O}=0$ for $m>2$, as each commutator decreases the differential order by 1. We also rescale $k$ by $k_a\rightarrow k_a\tau^{-1/2}$, so $\mathcal{L}_\phi\rightarrow\mathcal{L}_\phi\tau^{-1/2}$. Hence we have
\begin{equation}\label{Fouriershk2b}
\begin{aligned}
K&=\int \frac{d^4k}{(2\pi)^4\tau^2}e^\phi\exp\left[i\sum_{m=0}^{\infty}\frac{(-1)^m\tau^{1-m/2}}{m!}(\mathcal{L}_\phi)^m\mathcal{O}\right]E^{-1}y^\mu y_{\mu}y_{\dot{\mu}}y^{\dot{\mu}}\\
&=\int \frac{d^4k}{(2\pi)^4\tau^2}e^\phi\exp\left[i\left(\tau\mathcal{O}-\tau^{1/2}\mathcal{L}_\phi\mathcal{O}+\frac{\mathcal{L}_\phi{}^2}{2}\mathcal{O}\right)\right]E^{-1}y^\mu y_{\mu}y_{\dot{\mu}}y^{\dot{\mu}}.
\end{aligned}
\end{equation}
Comparing \eqref{Fouriershk2b} and \eqref{bnHKexpand}, we see that the coincidence limit of the heat kernel coefficients $[b_n]$, is given by
\begin{equation}\label{Fourierbn}
[b_n]=\frac{n!}{i^{n-1}}\left.\int \frac{d^4k}{\pi^2}\exp\left[i\left(\tau\mathcal{O}-\tau^{1/2}\mathcal{L}_\phi\mathcal{O}+\frac{\mathcal{L}_\phi{}^2}{2}\mathcal{O}\right)\right]E^{-1}(y^\mu)^2(y_{\dot{\mu}})^2\right|_{n, y^M\rightarrow 0},
\end{equation}
where $|_n$ means extracting out the coefficients of $\tau^n$. 

Note that such a formula is applicable not only for a superspace, but this can be generalized to \emph{any} space, one just needs to replace $E^{-1}(y^\mu)^2(y_{\dot{\mu}})^2$ by the appropriate counterpart. For example, for the chiral subspace we change $E^{-1}(y^\mu)^2(y_{\dot{\mu}})^2$ into the chiral version $\mathcal{E}^{-1}(\hat{y}^\mu)^2$ as in \cite{McArthur84}. Thus all the results here can be readily applied to the case of chiral superfields.

\subsection{Evaluation of Heat Kernel Coefficients via Power Series Expansion}\label{DilSYM2.2}
Roughly speaking in the coincidence limit, the effect of $\mathcal{L}_\phi$ is that it substitutes any bosonic covariant derivatives $\nabla_a$ that appear in the operator by $\nabla_a\rightarrow -ik_a$. Hence the term $\exp(i\mathcal{L}_\phi{}^2\mathcal{O}/2)$ will become $\exp(-i\psi k^ak_a)$ where $\psi$ is the coefficient of the d'Alembertian: $\mathcal{O}=\psi\Box+\cdots$. This provides the convergence for the $k_a$-integral in \eqref{Fourierbn} upon Wick rotation. Moreover, this term is independent of $\tau$, thus in calculating heat kernel coefficients, we shall isolate this term from the $\tau$ dependent piece in \eqref{Fourierbn}. One way to achieve this is to use the \emph{Baker-Hausdorff formula}, this was the approach used in \cite{McArthur84}. Here instead we shall expand the exponential differently as in \cite{Fisher85}, by a Dyson series type of expansion, which relies on the identity:
\begin{equation}\label{eDysonidentity}
\begin{aligned}
e^{A+B}=&e^A+\int_0^1d\alpha_1\,e^{\alpha_1A}Be^{(1-\alpha_1)A}\\
&\,+\int_0^1\int_0^{1-\alpha_1}d\alpha_1\,d\alpha_2 \,e^{\alpha_1A}Be^{\alpha_2A}Be^{(1-\alpha_1-\alpha_2)A}+\cdots.
\end{aligned}
\end{equation}
Borrowing the notation in \cite{Iochum17}, let us for convenience define the symbol
\begin{equation}\label{deffsymbol}
f^{(A)}_l[B_1\otimes\cdots\otimes B_l]=\int d\alpha\,e^{\alpha_1A}B_1e^{\alpha_2A}B_2\cdots B_l e^{(1-\alpha_1-\alpha_2-\cdots-\alpha_l)A},
\end{equation}
here the integration is understood to be the one in \eqref{eDysonidentity}:
\begin{equation*}
\int d\alpha=\int_0^1\int_0^{1-\alpha_1}\cdots\int_0^{1-\alpha_1-\cdots-\alpha_{l-1}} d\alpha_1\cdots d\alpha_l.
\end{equation*}
We shall call $l$ the "order" in the Dyson expansion. We can rewrite the identity in a simpler form:
\begin{equation}\label{eDysonidentity2}
e^{A+B}=e^A+f^{(A)}_1[B]+f^{(A)}_2[B\otimes B]+f^{(A)}_3[B\otimes B\otimes B]+\cdots.
\end{equation}

Here we should choose $A=i\mathcal{L}_\phi{}^2\mathcal{O}/2$ and $B=i\tau\mathcal{O}-i\tau^{1/2}\mathcal{L}_\phi\mathcal{O}$, and then apply the identity to expand the exponential in \eqref{Fourierbn}. All the $\tau$ dependence is now in the $B$ part, and it is easy to count the powers of $\tau$. For each $B$ in \eqref{eDysonidentity2}, we can choose either the term with $\mathcal{O}$ or the one with $\mathcal{L}_\phi\mathcal{O}$, this will result in different powers in $\tau$, and thus will ultimately contribute to different $[b_n]$. 

Let us sort the terms in \eqref{eDysonidentity2} by the powers in $\tau$. We might encounter terms proportional to half-integer powers in $\tau$. For instance we get a term with $\tau^{1/2}$ by choosing $\mathcal{L}_\phi\mathcal{O}$ in the first order expansion, then we get $\tau^{3/2}$ by choosing one copy of $\mathcal{L}_\phi\mathcal{O}$ and one copy of $\mathcal{O}$, or three copies of $\mathcal{L}_\phi\mathcal{O}$, and so on. However all these terms are odd under $k_a\rightarrow -k_a$, and thus they vanish after $k$-integration and will not be contributing to the heat kernel, as a result it suffices to keep only the terms with integer power. Now from \eqref{Fourierbn}, we have:
\begin{equation}\label{generalb0}
[b_0]=i\left.\int\frac{d^4k}{\pi^2}e^AE^{-1}(y^\mu)^2(y_{\dot{\mu}})^2\right|=0,
\end{equation}
where $|$ means taking the coincidence limit: $y^M\rightarrow 0$. $[b_0]=0$ is expected from supersymmetry, here it is due to the fact that $(y^\mu)^2(y_{\dot{\mu}})^2|=0$. In general, to get a non-zero result, one has to annihilate the term $(y^\mu)^2(y_{\dot{\mu}})^2$ by having covariant derivatives act on it, so it becomes non-vanishing in the coincidence limit. 

Next we have $[b_1]$, corresponding to the $\tau^1$ term. The result is:
\begin{equation}\label{generalb1}
[b_1]=\left.\int\frac{d^4k}{\pi^2}(if_1[\mathcal{O}]-f_2[\mathcal{L}_\phi\mathcal{O}\otimes\mathcal{L}_\phi\mathcal{O}])E^{-1}(y^\mu)^2(y_{\dot{\mu}})^2\right|.
\end{equation}
We see that the appearance of the operator $\mathcal{O}$ may lead to the annihilation of $(y^\mu)^2(y_{\dot{\mu}})^2$, as we might get derivative terms after the $k$-integration, and thus the result can be non-zero. Then for $[b_2]$, it is given by
\begin{equation}\label{generalb2}
\begin{aligned}
\left[b_2\right]=&2i\int\frac{d^4k}{\pi^2}(f_2[\mathcal{O}\otimes\mathcal{O}]+if_3[\mathcal{O}\otimes\mathcal{L}_\phi\mathcal{O}\otimes\mathcal{L}_\phi\mathcal{O}]\\
&+if_3[\mathcal{L}_\phi\mathcal{O}\otimes\mathcal{O}\otimes\mathcal{L}_\phi\mathcal{O}]
+if_3[\mathcal{L}_\phi\mathcal{O}\otimes\mathcal{L}_\phi\mathcal{O}\otimes\mathcal{O}]\\
&-f_4[\mathcal{L}_\phi\mathcal{O}\otimes\mathcal{L}_\phi\mathcal{O}\otimes\mathcal{L}_\phi\mathcal{O}\otimes\mathcal{L}_\phi\mathcal{O}])E^{-1}(y^\mu)^2(y_{\dot{\mu}})^2\left.\right|.
\end{aligned}
\end{equation}

We might continue and in theory one can express $[b_n]$ this way for any $n$. Similar to \eqref{Fourierbn}, by replacing $E^{-1}(y^\mu)^2(y_{\dot{\mu}})^2$ we may generalize the results here to other types of space, even a non-supersymmetric one. In that case we simply insert $E^{-1}$ without the fermionic coordinates.

To actually compute the coefficients, one has to perform Fourier integration of the functional $f_k$, at least in the coincidence limit. In other words, We have to compute the coincidence limit of $\int \frac{d^4k}{\pi^2}\,f_l[B_1\otimes\cdots B_l]$. The way to do so is to group the factors of exponentials in \eqref{deffsymbol}, by commuting the exponentials past the factors of $B$ using the identity \eqref{econjugateidentity}, for instance
\begin{equation}\label{fcommutation1}
\begin{aligned}
f_2[B_1\otimes B_2]&=\int d\alpha\,e^{\alpha_1A}B_1e^{\alpha_2A}B_2e^{(1-\alpha_1-\alpha_2)A}\\
&=\int d\alpha\,e^{\alpha_1A}B_1e^{\alpha_2A}e^{(1-\alpha_1-\alpha_2)A}e^{\mathcal{L}_{-(1-\alpha_1-\alpha_2)A}}B_2\\
&=\int d\alpha\,e^{A}e^{\mathcal{L}_{-(1-\alpha_1)A}}B_1e^{\mathcal{L}_{-(1-\alpha_1-\alpha_2)A}}B_2\\
&=\int d\alpha\,e^{A}\sum_{m,n=0}^{\infty}C_{m,n}(\alpha)(\mathcal{L}_A)^mB_1(\mathcal{L}_A)^nB_2,
\end{aligned}
\end{equation}
where $C_{m,n}(\alpha)=(-1)^{m+n}(1-\alpha_1)^m(1-\alpha_1-\alpha_2)^n/m!n!$. The summation is actually finite as $A$ contains no derivatives and thus $\mathcal{L}_A$ always decreases the differential order by 1. The $\alpha$ integral can be easily performed as only $C_{m,n}$ depends on $\alpha$ and it is just an elementary integral. This will give us a constant, say $D_{m,n}$. Now only the $k$ integral remains, so in the coincidence limit we will have expressions like 
\begin{equation}\label{fcommutation2}
\left.\int \frac{d^4k}{\pi^2}f_2[B_1\otimes B_2]\right|=\sum_{m,n=0}^{\infty}\int \frac{d^4k}{\pi^2}e^{-i\psi k^2}D_{m,n}(-ik^2)^{m+n}[(\mathcal{L}_\psi)^mB_1(\mathcal{L}_\psi)^nB_2],
\end{equation}
where $[(\mathcal{L}_\psi)^mB_1(\mathcal{L}_\psi)^nB_2]$ is the coincidence limit of the operator. We can rewrite this as:
\begin{equation}\label{fcommutation2b}
\left.\int \frac{d^4k}{\pi^2}f_2[B_1\otimes B_2]\right|=\int \frac{d^4k}{\pi^2}e^{-i\psi k^2}(F+G^{ab}k_ak_b+H^{abcd}k_ak_bk_ck_d+\cdots),
\end{equation}
for some operators $F$, $G^{ab}$, $H^{abcd}$, and so on. These $k$-integrals can be computed, and some simple results can be found in the literature. Finally we will get some local operator after integration, and this can be used to compute the heat kernel coefficients by acting it on $E^{-1}(y^\mu)^2(y_{\dot{\mu}})^2$. A more concrete example is shown in the appendix. To conclude, we have successfully demonstrated how to perform the relevant Fourier integrals to obtain $[b_n]$.

\section{First Three Heat Kernel Coefficients of a General Second Order Operator}\label{DilSYM3}
We have presented a method to calculate the super heat kernel coefficients, up to any order in principle. We are now going to derive a general formula for the first three heat kernel coefficients, for an arbitrary second order differential operator $\mathcal{O}$. These three coefficients will be crucial for studying the one-loop divergence of the corresponding theory. We shall restrict ourselves to the case of conformal supergravity, but one can readily apply the result to other types of supersymmetry theory with a different superspace, with only minor modifications required.

To start with, $[b_0]=0$ as required by supersymmetry. For $[b_1]$, from the general expression \eqref{generalb1} we have to find the coincidence limit of
\begin{equation*}
i\int\frac{d^4k}{\pi^2}f_1[\mathcal{O}]E^{-1}(y^\mu)^2(y_{\dot{\mu}})^2
\end{equation*}
and
\begin{equation*}
-\int\frac{d^4k}{\pi^2}f_2[\mathcal{L}_\phi\mathcal{O}\otimes\mathcal{L}_\phi\mathcal{O}]E^{-1}(y^\mu)^2(y_{\dot{\mu}})^2.
\end{equation*}
Let us start with the first one. In order to have a non-zero limit, we have to annihilate the factor $(y^\mu)^2(y_{\dot{\mu}})^2$ using covariant derivatives. In particular we need to find the terms containing four spinor derivatives, two dotted and two undotted, in the operator expression $\int\frac{d^4k}{\pi^2}f_1[\mathcal{O}]$. Now we have
\begin{equation}\label{Dilf1}
\begin{aligned}
f_1[\mathcal{O}]&=\int_0^1d\alpha_1\,e^{\alpha_1A}\mathcal{O}e^{(1-\alpha_1)A}\\
&=\int_0^1d\alpha_1\,e^{A}e^{\mathcal{L}_{-(1-\alpha_1)A}}\mathcal{O}.\\
\end{aligned}
\end{equation}
As $A$ is a constant and $\mathcal{O}$ is a second order differential operator,  $e^{\mathcal{L}_{-(1-\alpha_1)A}}\mathcal{O}$ is also a second order operator. Hence there cannot be any terms with four derivatives, and thus we conclude
\begin{equation*}
i\int\frac{d^4k}{\pi^2}f_1[\mathcal{O}]E^{-1}(y^\mu)^2(y_{\dot{\mu}})^2\rightarrow 0.
\end{equation*}
The same argument shows that
\begin{equation} \label{Dilb1f2}
\begin{aligned}
f_2[\mathcal{L}_\phi\mathcal{O}\otimes\mathcal{L}_\phi\mathcal{O}]&=
\int d\alpha\,e^{\alpha_1A}\mathcal{L}_\phi\mathcal{O}e^{\alpha_2A}\mathcal{L}_\phi\mathcal{O}e^{(1-\alpha_1-\alpha_2)A}\\
&=\int d\alpha\,e^{A}e^{\mathcal{L}_{-(1-\alpha_1)A}}\mathcal{L}_\phi\mathcal{O}e^{\mathcal{L}_{-(1-\alpha_1-\alpha_2)A}}\mathcal{L}_\phi\mathcal{O},\\
\end{aligned}
\end{equation}
which is of second order as $\mathcal{L}_\phi\mathcal{O}$ is a first order operator, thus it cannot contribute to $[b_1]$. Hence, the second heat kernel coefficient vanishes:
\begin{equation}\label{Dilb1}
[b_1]=0.
\end{equation}

\subsection{Calculation of $[b_2]$}\label{DilSYM3.1}
The next coefficient will be $[b_2]$, which is actually the first non-trivial one. From \eqref{generalb2} we have a handful of terms that will contribute, the first being
\begin{equation*}
2i\left.\int\frac{d^4k}{\pi^2}f_2[\mathcal{O}\otimes\mathcal{O}]E^{-1}(y^\mu)^2(y_{\dot{\mu}})^2\right|.
\end{equation*}
We need to extract the four spinor derivative terms in $f_2[\mathcal{O}\otimes\mathcal{O}]$, as in the case of $[b_1]$ there is a term
\begin{equation*}
\begin{aligned}
& \int d\alpha\,e^{\mathcal{L}_{-(1-\alpha_1)A}}\mathcal{O}e^{\mathcal{L}_{-(1-\alpha_1-\alpha_2)A}}\mathcal{O}\\
=&\sum_{m,n=0}^\infty \int_0^1\int_0^{1-\alpha_1} d\alpha_1d\alpha_2\,\frac{(1-\alpha_1)^m(1-\alpha_1-\alpha_2)^n}{m!\,n!}(\mathcal{L}_{-A})^m\mathcal{O}(\mathcal{L}_{-A})^n\mathcal{O}.
\end{aligned}
\end{equation*}
To get four derivatives, it is necessary to have $m=n=0$, as any commutators acting on $\mathcal{O}$ will lower the differential order. This also implies that only the part of $\mathcal{O}$ that contains only two spinor derivatives will contribute. Thus
\begin{equation}\label{Dilb2f2}
\begin{aligned}
\int\frac{d^4k}{\pi^2}f_2[\mathcal{O}\otimes\mathcal{O}]&\approx\int\frac{d^4k}{\pi^2}e^A\int_0^1\int_0^{1-\alpha_1} d\alpha_1d\alpha_2\mathcal{O}^2\\
&\rightarrow\int\frac{d^4k}{2\pi^2}e^{-i\psi k^2}\mathcal{O}^2
=-\frac{i}{2}\psi^{-2}\mathcal{O}^2.
\end{aligned}
\end{equation}
Here $\approx$ means equal up to terms with fewer spinor derivatives, which will have no significance, and we will not distinguish $\approx$ and $=$ in the following.

In general, one can write the part of $\mathcal{O}$ quadratic in spinor derivatives as
\begin{equation}\label{DilOquad}
\mathcal{O}=\psi F\nabla^2+\psi \bar{F}\bar\nabla^2+\psi V^{\alpha\dot{\alpha}}[\nabla_{\dot{\alpha}},\nabla_\alpha]+\cdots,
\end{equation}
where $F$, $\bar{F}$ and $V^{\alpha\dot{\alpha}}$ are some arbitrary fields. Note that we have isolated the factor $\psi$ which will make the final answer simple. We just have to calculate from this the terms with four spinor derivatives in $\mathcal{O}^2$, which is easily seen:
\begin{equation}\label{O24thorder}
\mathcal{O}^2=\psi^2(2F\bar{F}+V^{\alpha\dot{\alpha}}V_{\alpha\dot{\alpha}})\nabla^2\bar{\nabla}^2.
\end{equation}
Recall that we are omitting all terms with less than four spinor derivatives. To derive the equation, we have used some identities like $\nabla^2\bar{\nabla}^2=\bar{\nabla}^2\nabla^2+\cdots$,  $[\nabla_{\dot{\alpha}},\nabla_\alpha][\nabla_{\dot{\beta}},\nabla_\beta]=-4\nabla_{\dot{\alpha}}\nabla_{\dot{\beta}}\nabla_\alpha\nabla_\beta+\cdots$, and also 
\begin{equation}\label{nabla2simplify}
\nabla_{\alpha}\nabla_{\beta}=\frac{1}{2}\epsilon_{\alpha\beta}\nabla^2
\end{equation}
together with its conjugate, which can be proved from the fact that in conformal supergravity $\{\nabla_{\alpha},\nabla_{\beta}\}=0$.\footnote{In general for a different superspace, the formula \eqref{nabla2simplify} will still be true if we discard terms with fewer derivatives.} Combining \eqref{Dilb2f2} and \eqref{O24thorder}, we arrive at the $[b_2]$ contribution:
\begin{equation}\label{Dilb2contri1}
[b_2]\ni 2i\left.\int\frac{d^4k}{\pi^2}f_2[\mathcal{O}\otimes\mathcal{O}]E^{-1}(y^\mu)^2(y_{\dot{\mu}})^2\right|=16V^{\alpha\dot{\alpha}}V_{\alpha\dot{\alpha}}+32F\bar{F}.
\end{equation}

The next contribution will be 
\begin{equation*}
-2\left.\int\frac{d^4k}{\pi^2}f_3[\mathcal{O}\otimes\mathcal{L}_\phi\mathcal{O}\otimes\mathcal{L}_\phi\mathcal{O}]E^{-1}(y^\mu)^2(y_{\dot{\mu}})^2\right|,
\end{equation*}
in which we will encounter the expression
\begin{equation*}
\int d\alpha\,e^{\mathcal{L}_{-(1-\alpha_1)A}}\mathcal{O}e^{\mathcal{L}_{-(1-\alpha_1-\alpha_2)A}}\mathcal{L}_{\phi}\mathcal{O}e^{\mathcal{L}_{-(1-\alpha_1-\alpha_2-\alpha_3)}}\mathcal{L}_{\phi}\mathcal{O}.
\end{equation*}
As $\mathcal{L}_{\phi}\mathcal{O}$ is of first order, this is at most fourth order in derivatives. Thus we have to again choose the part of $\mathcal{O}$ with two spinor derivatives and the spinor derivative part of $\mathcal{L}_{\phi}\mathcal{O}$ in order to have a non-zero result. We also replace all the exponentials by 1, as other terms in the Taylor expansion will contain commutators, which will lower the differential order. The integration over $\alpha$ is now trivial, and gives $1/3!=1/6$. What remains is to find the fourth order spinor derivative part of 
\begin{equation*}
-\frac{1}{3}\int\frac{d^4k}{\pi^2}e^A\mathcal{O}(\mathcal{L}_\phi\mathcal{O})^2,
\end{equation*}
at the coincidence limit. If $\mathcal{O}$ contains the term
\begin{equation}\label{DilOquad2}
\mathcal{O}=\psi X^{a\alpha}\{\nabla_a,\nabla_{\alpha}\}-\psi \bar{X}^{a\dot{\alpha}}\{\nabla_a,\nabla_{\dot\alpha}\}+\cdots,
\end{equation}
then in the coincidence limit, we get
\begin{equation}\label{DilLOderi}
[\mathcal{L}_{\phi}\mathcal{O}]=-2i\psi X^{a\alpha}k_a\nabla_{\alpha}+2i\psi \bar{X}^{a\dot{\alpha}}k_a\nabla_{\dot\alpha}+\cdots,
\end{equation}
note that there are other terms in $\mathcal{O}$
that produce a spinor derivative in $\mathcal{L}_{\phi}\mathcal{O}$. For example $\psi F\nabla^2$ gives rise to the contribution $-2\psi\nabla^\alpha\phi\nabla_{\alpha}$, but it vanishes in the coincidence limit. The desired quartic spinor derivative term is then
\begin{equation}\label{OLO24thorder}
[\mathcal{O}(\mathcal{L}_\phi\mathcal{O})^2]=2\psi^3(2X^{a\alpha}\bar{X}^{b\dot{\alpha}}V_{\alpha\dot{\alpha}}-\bar{F}X^{a\alpha}X^{b}{}_{\alpha}-F\bar{X}^a{}_{\dot{\alpha}}\bar{X}^{b\dot{\alpha}})k_ak_b\nabla^2\bar{\nabla}^2,
\end{equation}
which can be seen by calculating the part of $[(\mathcal{L}_{\phi}\mathcal{O})^2]$ with two spinor derivatives:
\begin{equation}\label{DilLO2quad}
[(\mathcal{L}_\phi\mathcal{O})^2]=2\psi^2k_ak_b(2X^{a\alpha}\bar{X}^{b\dot{\alpha}}[\nabla_{\dot\alpha},\nabla_{\alpha}]-X^{a\alpha}X^{b}{}_{\alpha}\nabla^2-\bar{X}^a{}_{\dot{\alpha}}\bar{X}^{b\dot{\alpha}}\bar\nabla^2).
\end{equation}
We use the following identity to integrate over $k$:
\begin{equation}\label{kabintidentity}
\begin{aligned}
\int\frac{d^4k}{\pi^2}e^{-i\psi k^2}k_ak_b&=\int\frac{d^4k}{4\pi^2}e^{-i\psi k^2}k^2\eta_{ab}\\
&=\frac{i}{4}\eta_{ab}\frac{\partial}{\partial\psi}\left(\int\frac{d^4k}{\pi^2}e^{-i\psi k^2}\right)\\
&=\frac{i}{4}\eta_{ab}\frac{d}{d\psi}(-i\psi^{-2})=-\frac{\eta_{ab}}{2}\psi^{-3}.
\end{aligned}
\end{equation}
Here in the first line, we have used the fact that the original integral is symmetric in $a$ and $b$, so the final expression must be proportional to $\eta_{ab}$. This leads to the final result
\begin{equation*}
\begin{aligned}
&-2\left.\int\frac{d^4k}{\pi^2}f_3[\mathcal{O}\otimes\mathcal{L}_\phi\mathcal{O}\otimes\mathcal{L}_\phi\mathcal{O}]E^{-1}(y^\mu)^2(y_{\dot{\mu}})^2\right|\\
=&\frac{1}{3}(32X^{a\alpha}\bar{X}_{a}{}^{\dot{\alpha}}V_{\alpha\dot{\alpha}}-16\bar{F}X^{a\alpha}X_{a\alpha}-16F\bar{X}_{a\dot{\alpha}}\bar{X}^{a\dot{\alpha}}).
\end{aligned}
\end{equation*}
We quickly realize that the term containing $f_3[\mathcal{L}_\phi\mathcal{O}\otimes\mathcal{O}\otimes\mathcal{L}_\phi\mathcal{O}]$ and also the one with $f_3[\mathcal{L}_\phi\mathcal{O}\otimes\mathcal{L}_\phi\mathcal{O}\otimes\mathcal{O}]$ will give the same result. Combining these three contributions we have
\begin{equation}\label{Dilb2contri2}
[b_2]\ni 32X^{a\alpha}\bar{X}_{a}{}^{\dot{\alpha}}V_{\alpha\dot{\alpha}}-16\bar{F}X^{a\alpha}X_{a\alpha}-16F\bar{X}_{a\dot{\alpha}}\bar{X}^{a\dot{\alpha}}.
\end{equation}

The one last term we need to deal with is
\begin{equation*}
-2i\left.\int\frac{d^4k}{\pi^2}f_4[\mathcal{L}_\phi\mathcal{O}\otimes\mathcal{L}_\phi\mathcal{O}\otimes\mathcal{L}_\phi\mathcal{O}\otimes\mathcal{L}_\phi\mathcal{O}]E^{-1}(y^\mu)^2(y_{\dot{\mu}})^2\right|.
\end{equation*}
With the same arguments as above, it suffices to isolate the fourth order spinor derivative term of
\begin{equation*}
\left.\int d\alpha (\mathcal{L}_\phi\mathcal{O})^4\right|=\frac{1}{4!}[(\mathcal{L}_\phi\mathcal{O})^4].
\end{equation*}
With the help of \eqref{DilLO2quad} we get
\begin{equation}\label{LO44thorder}
\frac{1}{4!}[(\mathcal{L}_\phi\mathcal{O})^4]=\frac{1}{3}\psi^4k_ak_bk_ck_d(X^{a\alpha}X^{b}{}_{\alpha}\bar{X}^{c}{}_{\dot{\alpha}}\bar{X}^{d\dot{\alpha}}+2X^{a\alpha}\bar{X}^{b\dot{\alpha}}X^{c}{}_{\alpha}\bar{X}^{d}{}_{\dot{\alpha}})\nabla^2\bar\nabla^2.
\end{equation}
For the $k$-integral, we need, using symmetry arguments, 
\begin{equation}\label{kabcdintidentity}
\begin{aligned}
\int\frac{d^4k}{\pi^2}e^{-i\psi k^2}k_ak_bk_ck_d&=\int\frac{d^4k}{\pi^2}e^{-i\psi k^2}\left[\frac{k^4}{4!}(\eta_{ab}\eta_{cd}+\eta_{ac}\eta_{bd}+\eta_{ad}\eta_{bc})\right]\\
&=-\frac{1}{24}(\eta_{ab}\eta_{cd}+\eta_{ac}\eta_{bd}+\eta_{ad}\eta_{bc})\frac{\partial^2}{\partial\psi^2}\left(\int\frac{d^4k}{\pi^2}e^{-i\psi k^2}\right)\\
&=\frac{i}{4}(\eta_{ab}\eta_{cd}+\eta_{ac}\eta_{bd}+\eta_{ad}\eta_{bc})\psi^{-4}.
\end{aligned}
\end{equation}
After some work, we obtain the final piece of the coefficient $[b_2]$:
\begin{equation}\label{Dilb2contri3}
[b_2]\ni 8X^{a\alpha}X_{a\alpha}\bar{X}_{b\dot{\alpha}}\bar{X}^{b\dot{\alpha}}+16X^{a\alpha}X_{b\alpha}\bar{X}_{a\dot{\alpha}}\bar{X}^{b\dot{\alpha}}.
\end{equation}

Combining all the results, we have the final answer: for a general second operator $\mathcal{O}$ with its quadratic part given by
\begin{equation}\label{generalO2ndorder}
\mathcal{O}^{(2)}=\psi(\Box+F\nabla^2+\bar{F}\bar\nabla^2+ V^{\alpha\dot{\alpha}}[\nabla_{\dot{\alpha}},\nabla_\alpha]+X^{a\alpha}\{\nabla_a,\nabla_{\alpha}\}-\bar{X}^{a\dot{\alpha}}\{\nabla_a,\nabla_{\dot\alpha}\}),
\end{equation}
and its third heat kernel coefficient $[b_2]$ is given by
\begin{equation}\label{generalO2b2}
\begin{aligned}
\left[b_2\right]=&\,32F\bar{F}+16V^{\alpha\dot{\alpha}}V_{\alpha\dot{\alpha}}+32X^{a\alpha}\bar{X}_{a}{}^{\dot{\alpha}}V_{\alpha\dot{\alpha}}-16\bar{F}X^{a\alpha}X_{a\alpha}\\
&\,-16F\bar{X}_{a\dot{\alpha}}\bar{X}^{a\dot{\alpha}}+8X^{a\alpha}X_{a\alpha}\bar{X}_{b\dot{\alpha}}\bar{X}^{b\dot{\alpha}}+16X^{a\alpha}X_{b\alpha}\bar{X}_{a\dot{\alpha}}\bar{X}^{b\dot{\alpha}}.
\end{aligned}
\end{equation}
As a consistency check, we can compare this with the calculation using the previously developed non-recursive method \cite{Leung19}. In the latter case, we have $\psi=1$, and we imposed $X^{a\alpha}=\bar{X}^{a\dot{\alpha}}=0$ as a constraint, it is clear that the two results agree in this special case. The expression shown in \eqref{generalO2b2} can be regarded as a more general result, covering the possibility in which $\mathcal{O}$ contains mixed derivative terms like $X^{a\alpha}\nabla_a\nabla_{\alpha}$.

The crucial result \eqref{generalO2b2} in fact holds for \emph{all} types of $N=1$ superspace, despite we have restricted ourselves to the case of conformal supergravity at the moment. Firstly this is because the expression is an algebraic one with no derivatives, thus it is independent of how the covariant derivatives are defined. Secondly, for other superspaces the covariant derivative algebra will be different and possibly be more complicated. But by carefully looking at our derivation we see that the commutation algebra plays no role in determining $[b_2]$. However, both reasonings will break down for other heat kernel coefficients so this feature is exclusive for $[b_2]$ only.

Note that this expression of $[b_2]$ is independent of $\psi$, which might be a surprise but this is merely due to how the functions $F$, $\bar{F}$, and such are defined. Indeed, if $\psi$ is a constant, we expect the heat kernel coefficients $[b_2]$ to be independent of $\psi$. This is because $[b_2]$ controls the logarithmic divergence of the corresponding theory which is scheme independent, thus it cannot depend on an overall pre-factor $\psi$, which can be absorbed by a simple rescaling.
In fact, in general a constant rescaling of $\psi$: $\psi\rightarrow\lambda\psi$, which is roughly equivalent to rescaling the whole operator $\mathcal{O}\rightarrow\lambda\mathcal{O}$, will incur a change in heat kernel coefficients: $[b_n]\rightarrow \lambda^{n-2}[b_n]$.\footnote{This can be easily seen from the operator definition of the super heat kernel.} For $n=2$, it is indeed independent of the transformation as desired.

We also note that we have set the coefficient $\psi$ to be a scalar, thus it naturally commutes with other coefficients like $F$. But the method presented here also applies if this is not the case, in particular when $\psi$ is matrix valued. Only a minor modification of \eqref{generalO2b2} is needed, we would have to insert various powers of $\psi$ in the appropriate places and we would have to take the trace at the end.

\subsection{Higher Order Heat Kernel Coefficients}\label{DilSYM3.2}
We shall briefly describe here some general features that will appear in the computation of higher order heat kernel coefficients $[b_n]$. From the general expression of $[b_n]$, equation \eqref{Fourierbn}, we will have contributions going from $n$-th order in the Dyson expansion, in particular one proportional to $f_n[\mathcal{O}\otimes\cdots\otimes\mathcal{O}]$, to a $2n$-th order term that depends on $f_{2n}[\mathcal{L}_\phi\mathcal{O}\otimes\cdots\otimes\mathcal{L}_\phi\mathcal{O}]$. All these operators are of $2n$-th differential order. 

As in the case of $[b_2]$, we need the terms that contain two undotted and two dotted spinor derivatives. For a general $n>2$, we see that instead of only the quadratic part, the linear and the constant part of $\mathcal{O}$ will also come into play, for instance in $f_n[\mathcal{O}\otimes\cdots\otimes\mathcal{O}]$, we will still have a sufficient number of spinor derivatives even if we choose the lower order part for some of the $\mathcal{O}$'s. Also, we see that the effect of $\exp(\mathcal{L}_{-(1-\cdots-\alpha_k)A})$ appearing in the $f_n$ functions is non-trivial, as opposed to the case of $[b_2]$. We have enough room to include these commutators that decrease the differential order, as we start with $2n$-th order and we only need fourth order. Thus we expect that there will be terms that depend on derivatives of $\psi$, up to the $(2n-4)$-th order.

After performing the $k$-integration, by mimicking the trick used above, we will have a differential operator that acts on $E^{-1}(y^\mu)^2(y_{\dot{\mu}})^2$, then we take the coincidence limit and obtain the heat kernel coefficients.
In general, any differential operators can be written in form
\begin{equation}\label{generaldiffop}
\mathcal{Q}\nabla^2\bar\nabla^2+\textrm{terms with fewer spinor derivatives,}
\end{equation}
where $\mathcal{Q}$ is some operator. For $[b_n]$, this part is of $(2n-4)$-th order. The term $\nabla^2\bar\nabla^2$ will annihilate $(y^\mu)^2(y_{\dot{\mu}})^2$ and gives a non-zero result. Then $\mathcal{Q}$ can act on $E^{-1}$, hence we need the coincidence limit of the $(2n-4)$-th order derivative of $E^{-1}$. This can be achieved by using the normal coordinate expansion. Either one can obtain the normal expansion of the vielbein \cite{McArthur84b}\cite{Grisaru98} and calculate the determinant, or one can use the iterative method as in \cite{Kuzenko09}. In fact in the context of conformal supergravity, due to the fact that 
\begin{equation*}
\{\nabla_{\alpha},\nabla_{\beta}\}=\{\nabla_{\dot{\alpha}},\nabla_{\dot{\beta}}\}=0,
\end{equation*}
$\mathcal{Q}=A+B^a\nabla_a+C\Box+\cdots$ is an operator that is constructed \emph{only} from the bosonic covariant derivative $\nabla_a$. Hence we only need the expansion in the $y^m$ direction. But for a general supergravity theory, the full normal expansion is required.

Also, in the calculation of obtaining the final operator as in \eqref{generaldiffop}, we often encounter higher order derivatives of $\phi=ik_a\delta^a{}_my^m$. One just need the equation $\nabla_A(y^m)=E_A{}^m$ and the normal coordinate expansion of the vielbein. For example the second order derivative of $\phi$ will involve the torsion tensor $T_{AB}{}^C$. Hence with the help of the normal coordinate expansion, one can theoretically calculate the heat kernel coefficients up to any order.

As a final remark, we so far focused on operators in the full superspace, but the machinery presented here applies to the case of chiral fields by applying the same method in the chiral subspace. In fact, it should be possible, at least in theory, to generalize such a method to \emph{any} superspace, not only in four dimensions. However the possibility of such a generalization, while interesting, will not be discussed here. 

\section{One-loop Divergence of SYM with a Dilaton}\label{DilSYM4}
As an application, let us determine the first three heat kernel coefficients for the super Yang-Mills model coupled to a dilaton, with our operator of interest being $\mathcal{O}_{V,S}$ as derived in \eqref{DilYMOV}. As discussed in the case of a trivial gauge kinetic function \cite{Leung19}, in conformal supergravity, the operator governing the one-loop effective action for a vector multiplet is actually not invariant under dilation: $[D,\mathcal{O}_{V,S}]=2\mathcal{O}_{V,S}$. Hence complications arise when we have to exponentiate the operator to define the heat kernel, as the exponential will not be an invariant object. One method to resolve this is to demote the $D$-symmetry, no longer treating it as gauged temporarily, and checking $D$-invariance at the end. Instead the route we take here in order to regulate the symmetry is to make use of the \emph{compensator} $X$, which was already introduced for Yang-Mills gauge-fixing and it satisfies $DX=2X$. We then have $X^{-1/2}\mathcal{O}_{V,S}X^{-1/2}$ being $D$-invariant thus we can proceed normally.

In fact, considering $X^{-1/2}\mathcal{O}_{V,S}X^{-1/2}$ is equivalent to redefining the quanta of the vector multiplet by $V'=X^{1/2}V$. It is clear that the quadratic action of $V'$ is 
\begin{equation*}
\begin{aligned}
S^{(2)}&=\frac{1}{2}\tr\int d^8z \,E V\mathcal{O}_{V,S}V\\
&=\frac{1}{2}\tr\int d^8z \,E V'(X^{-1/2}\mathcal{O}_{V,S}X^{-1/2})V',
\end{aligned}
\end{equation*}
hence the reason for picking this particular combination. As a remark, the regulation scheme described here is just one way to proceed. Different schemes are equivalent in the sense that they will give the same result on-shell.

With this technicality settled, we shall now consider the heat kernel coefficients of the operator
\begin{equation}\label{DilhkO}
\mathcal{O}=X^{-1/2}\mathcal{O}_{V,S}X^{-1/2}+X^{-1}m^2.
\end{equation}
Here we also introduced a potential mass matrix term $m^2$ for the vector multiplet, which comes from the background field expansion of the K\"ahler potential $K$ if some chiral fields involved carry non-trivial Yang-Mills charges. However, we have seen that such a mass term will not contribute to the first three coefficients, but only to the higher order ones.

\subsection{First Three Heat Kernel Coefficients of the Vector Multiplet}\label{DilSYM4.1}
We already know the first two coefficients are zero: $[b_0]=[b_1]=0$. To obtain $[b_2]$, we need to find the various objects appearing in the general formula \eqref{generalO2b2}, which can be read off from the quadratic part of $\mathcal{O}$.
This is just a straightforward task from the derived form of $\mathcal{O}_{V,S}$, shown in \eqref{DilYMOV}. We have
\begin{equation}\label{DilSYMquad}
\begin{aligned}
\psi=&\,X^{-1}(S+\bar{S}),\\
F=\frac{1}{16}\left(2U_{\dot{\alpha}}U^{\dot{\alpha}}+\frac{\bar\nabla^2\bar{S}}{S+\bar{S}}\right),& \quad \bar{F}=\frac{1}{16}\left(2U^{\alpha}U_{\alpha}+\frac{\nabla^2S}{S+\bar{S}}\right),\\
V^{\alpha\dot{\alpha}}=\frac{1}{2}G^{\alpha\dot{\alpha}}+\frac{1}{2}U^\alpha U^{\dot{\alpha}}&-\frac{\nabla^{\alpha}SU^{\dot{\alpha}}+U^\alpha\nabla^{\dot{\alpha}}\bar{S}}{16(S+\bar{S})},\\
X^{a\alpha}=\frac{i}{4}\left(U_{\dot{\alpha}}-\frac{\nabla_{\dot{\alpha}}\bar{S}}{S+\bar{S}}\right)(\bar\sigma^{a})^{\dot{\alpha}\alpha},& \quad \bar{X}_{a\dot{\alpha}}=\frac{i}{4}\left(U^\alpha-\frac{\nabla^{\alpha} S}{S+\bar{S}}\right)(\sigma_a)_{\alpha\dot{\alpha}}.
\end{aligned}
\end{equation}

We can directly use \eqref{generalO2b2} to obtain $[b_2]$, but it is immediately seen that the algebra involved is getting quite tedious. To simplify the calculation, we shall employ a strategy as follows: We choose the special conformal gauge $U^\alpha=U_{\dot{\alpha}}=0$ which breaks the $K_A$ symmetry. This eliminates most of the terms in \eqref{DilSYMquad}, and we compute $[b_2]$ in this particular case. We restore the $K_A$ invariance by demanding the actual pre-gauge-fixed expression is \emph{conformal primary}, in the sense that the expression has to be annihilated by $K_A$. This can be achieved by adding correction terms that depend on $U^\alpha$ and $U_{\dot{\alpha}}$.

With this specific choice of gauge, the relevant quantities become simple:
\begin{equation}\label{DilSYMquad2}
\begin{aligned}
F'=\frac{\bar\nabla^2\bar{S}}{16(S+\bar{S})},& \quad \bar{F}'=\frac{\nabla^2S}{16(S+\bar{S})},\\
V'{}^{\alpha\dot{\alpha}}&=\frac{1}{2}G^{\alpha\dot{\alpha}},\\
X'{}^{a\alpha}=-\frac{i}{4}\frac{\nabla_{\dot{\alpha}}\bar{S}}{S+\bar{S}}(\bar\sigma^{a})^{\dot{\alpha}\alpha},& \quad \bar{X}'_{a\dot{\alpha}}=-\frac{i}{4}\frac{\nabla^{\alpha} S}{S+\bar{S}}(\sigma_a)_{\alpha\dot{\alpha}}.
\end{aligned}
\end{equation}
Let us calculate some of the expressions that will be useful:
\begin{equation}\label{X2contraction}
\begin{aligned}
X'{}^{a\alpha}X'{}_{a\alpha}&=-\frac{1}{16(S+\bar{S})^2}\nabla_{\dot{\alpha}}\bar{S}(\bar\sigma^{a})^{\dot{\alpha}\alpha}\nabla_{\dot{\beta}}\bar{S}(\bar\sigma^{a}\epsilon)^{\dot{\beta}}{}_{\alpha}\\
&=-\frac{1}{16(S+\bar{S})^2}\nabla_{\dot{\alpha}}\bar{S}\nabla_{\dot{\beta}}\bar{S}(-2\delta^{\alpha}{}_{\alpha}\epsilon^{\dot{\alpha}\dot{\beta}})\\
&=-\frac{1}{4(S+\bar{S})^2}\nabla_{\dot{\alpha}}\bar{S}\nabla^{\dot{\alpha}}\bar{S}.
\end{aligned}
\end{equation}
We have used the identity $(\bar\sigma^{a})^{\dot{\alpha}\alpha}(\bar\sigma_{a})^{\dot{\beta}\beta}=-2\epsilon^{\alpha\beta}\epsilon^{\dot{\alpha}\dot{\beta}}$, which can be found in for instance the appendix of \cite{BBG00}. Similarly we have
\begin{equation}\label{barX2contraction}
\bar{X}'_{a\dot\alpha}\bar{X}'{}^{a\dot\alpha}=-\frac{1}{4(S+\bar{S})^2}\nabla^{\alpha}S\nabla_{\alpha}S.
\end{equation}
We also have
\begin{equation}\label{XXbarcontraction}
\begin{aligned}
X'{}^{a\alpha}\bar{X}'{}_{a\dot{\alpha}}&=-\frac{1}{16(S+\bar{S})^2}\nabla_{\dot{\beta}}\bar{S}(\bar\sigma^{a})^{\dot{\beta}\alpha}\nabla^{\beta}S(\sigma_{a})_{\beta\dot{\alpha}}\\
&=-\frac{1}{16(S+\bar{S})^2}\nabla_{\dot{\beta}}\bar{S}\nabla^{\beta}S(-2\delta^{\dot{\beta}}{}_{\dot{\alpha}}\delta^{\alpha}{}_{\beta})\\
&=\frac{1}{8(S+\bar{S})^2}\nabla^{\alpha}S\nabla_{\dot{\alpha}}\bar{S}.
\end{aligned}
\end{equation}

With some algebra, we arrive at the expression, using the general result \eqref{generalO2b2}:
\begin{equation}\label{DilSYMgfb2}
\begin{aligned}
\left[b_2\right]'=\,&\frac{1}{8(S+\bar{S})^2}\bar\nabla^2\bar{S}\nabla^2S+4G^{\alpha\dot{\alpha}}G_{\alpha\dot{\alpha}}+\frac{2}{(S+\bar{S})^2}\nabla^{\alpha}SG_{\alpha\dot{\alpha}}\nabla^{\dot{\alpha}}\bar{S}\\
&+\frac{1}{4(S+\bar{S})^3}\nabla^2S\nabla_{\dot{\alpha}}\bar{S}\nabla^{\dot{\alpha}}\bar{S}+\frac{1}{4(S+\bar{S})^3}\bar\nabla^2\bar{S}\nabla^{\alpha}S\nabla_{\alpha}S\\
&+\frac{3}{4(S+\bar{S})^4}\nabla_{\dot{\alpha}}\bar{S}\nabla^{\dot{\alpha}}\bar{S}\nabla^{\alpha}S\nabla_{\alpha}S\\
=&\,-8G^aG_a+\frac{2}{(S+\bar{S})^2}\nabla^{\alpha}SG_{\alpha\dot{\alpha}}\nabla^{\dot{\alpha}}\bar{S}\\
&+\frac{1}{8(S+\bar{S})^2}\left(\bar\nabla^2\bar{S}+2\frac{\nabla_{\dot\alpha}\bar{S}\nabla^{\dot{\alpha}}\bar{S}}{S+\bar{S}}\right)\left(\nabla^2S+2\frac{\nabla^{\alpha}S\nabla_{\alpha}S}{S+\bar{S}}\right)\\
&+\frac{1}{4(S+\bar{S})^4}\nabla_{\dot{\alpha}}\bar{S}\nabla^{\dot{\alpha}}\bar{S}\nabla^{\alpha}S\nabla_{\alpha}S.
\end{aligned}
\end{equation}

We shall now consider the correct terms that have to be added in order to recover the special conformal symmetry, $K_A$. This is equivalent to saying that the final expression has to be annihilated by the operators $S^\alpha$ and $\bar{S}_{\dot{\alpha}}$. Among all the fields appearing in \eqref{DilSYMgfb2}, it is not hard to show that only the second derivatives $\bar\nabla^2\bar{S}$ and $\nabla^2S$ are not conformal primary. Simple calculation shows 
\begin{equation}\label{DDSnonprimary}
\begin{aligned}
S^\alpha\bar\nabla^2\bar{S}=0,& \quad \bar{S}_{\dot\alpha}\bar\nabla^2\bar{S}=8\nabla_{\dot\alpha}\bar{S},\\
S^\alpha\nabla^2S=8\nabla^{\alpha}S,& \quad \bar{S}_{\dot\alpha}\nabla^2S=0.
\end{aligned}
\end{equation}
We have to cancel the non-zero charge by introducing corrections that depend on $U^\alpha=X^{-1}\nabla^{\alpha}X$ and $U_{\dot{\alpha}}=X^{-1}\nabla_{\dot{\alpha}}X$. Some algebra leads to
\begin{equation}\label{UAnonprimary}
\begin{aligned}
S^{\beta}U^\alpha=-4\epsilon^{\beta\alpha},& \quad \bar{S}_{\dot\beta}U^\alpha=0,\\
S^{\beta}U_{\dot{\alpha}}=0,& \quad \bar{S}_{\dot\beta}U_{\dot{\alpha}}=-4\epsilon_{\dot{\beta}\dot{\alpha}}.
\end{aligned}
\end{equation}

Using \eqref{DDSnonprimary} and \eqref{UAnonprimary}, we see that the combinations
$\nabla^2S+2U^\alpha\nabla_{\alpha}S$ and $\bar\nabla^2\bar{S}+2U_{\dot{\alpha}}\nabla^{\dot{\alpha}}\bar{S}$ are then conformal primary.\footnote{Alternatively, one can show $\nabla^2S+2U^\alpha\nabla_{\alpha}S=X^{-1}(\nabla^2+8R)(XV)$ which is a conformal primary expression.} Hence one just has to make the substitutions $\nabla^2S\rightarrow\nabla^2S+2U^\alpha\nabla_{\alpha}S$ and the conjugate $\bar\nabla^2\bar{S}\rightarrow\bar\nabla^2\bar{S}+2U_{\dot{\alpha}}\nabla^{\dot{\alpha}}\bar{S}$ in \eqref{DilSYMgfb2}, and the resulting expression will be conformal primary.

To see that there are no more terms to be added to $[b_2]'$, note that the actual $[b_2]$ before gauge fixing must be constructed from the objects $G^{\alpha\dot{\alpha}}$, $S+\bar{S}$, $\nabla^{\alpha}S$, $\nabla_{\dot{\alpha}}\bar{S}$, $U^\alpha$, $U_{\dot{\alpha}}$, $\nabla^2S$ and $\bar\nabla^2\bar{S}$. It is a straightforward verification that no other correction terms, which must vanish for the gauge choice $U^\alpha=U_{\dot{\alpha}}=0$ while also being conformal primary, can be introduced in \eqref{DilSYMgfb2}. Hence the final conformal invariant expression is given by
\begin{equation}\label{DilSYMtrueb2}
\begin{aligned}
\left[b_2\right]=\,&-8G^aG_a+\frac{2}{(S+\bar{S})^2}\nabla^{\alpha}SG_{\alpha\dot{\alpha}}\nabla^{\dot{\alpha}}\bar{S}\\
&+\frac{1}{8(S+\bar{S})^2}\bar\Sigma\Sigma+\frac{1}{4(S+\bar{S})^4}\nabla_{\dot{\alpha}}\bar{S}\nabla^{\dot{\alpha}}\bar{S}\nabla^{\alpha}S\nabla_{\alpha}S,
\end{aligned}
\end{equation}
where 
\begin{equation}\label{Sigmadef}
\begin{aligned}
&\Sigma=\nabla^2S+2U^\alpha\nabla_{\alpha}S+2\frac{\nabla^{\alpha}S\nabla_{\alpha}S}{S+\bar{S}},\\
&\bar{\Sigma}=\bar\nabla^2\bar{S}+2U_{\dot{\alpha}}\nabla^{\dot{\alpha}}\bar{S}+2\frac{\nabla_{\dot\alpha}\bar{S}\nabla^{\dot{\alpha}}\bar{S}}{S+\bar{S}}.
\end{aligned}
\end{equation}
As a remark, to obtain the corresponding expression in the $U(1)$ or Poincar\'e supergravity, one just has to choose the conformal gauge $U_A=0$, and replace the conformal covariant derivatives $\nabla_A$ by the covariant derivatives post-gauge-fixing, $\mathcal{D}_A$. More details on obtaining various supergravity theories via gauge-fixing the superconformal symmetry can be found in \cite{Butter10}.

\subsection{Ghost Contributions}\label{DilSYM4.2}
The derived $[b_2]$ in \eqref{DilSYMtrueb2} allows us to obtain the logarithmic divergence of SYM due to the vector multiplet. However we have to consider the ghost fields also in order to have the full divergence. Thus we turn to the one-loop divergences of the ghost action next. It is easily seen that the quadratic divergence is the same as the case without the dilaton, thus we will not discuss it here and we focus on the logarithmic divergence.

Let us start with the Faddeev-Popov ghost. As we are using the identical gauge-fixing functional as the constant coupling case, $f-\bar{\nabla}^2(XV)=\bar{f}-\nabla^2(XV)=0$, we have the same Faddeev-Popov ghost action. As a result, we also have the same induced logarithmic divergence, which is given by the expression \cite{Leung19}
\begin{equation}\label{FP1loopdivrecall}
\begin{aligned}
\Gamma^{\textrm{FP}}_{(1)\textrm{log}}=&\,\frac{\log{\Lambda^2}}{48\pi^2}S_{\chi}-\frac{\log{\Lambda^2}}{16\pi^2}[4R\bar{R}]_D\\
&-\frac{\log{\Lambda^2}}{32\pi^2}\left(\left[\left(\mathcal{W}^{\alpha}_{\textrm{YM}}+\frac{1}{3}X^{\alpha}\right)^2+\frac{2}{3}W^{\alpha\beta\gamma}W_{\gamma\beta\alpha}\right]_F+\textrm{h.c.}\right),
\end{aligned}
\end{equation}
where $S_{\chi}=[G^aG_a+2R\bar{R}]_D+\left(\left[\frac{1}{12}X^\alpha X_\alpha+\frac{1}{2}W^{\alpha\beta\gamma}W_{\gamma\beta\alpha}\right]_F+\textrm{h.c.}\right)$ is a topological invariant. One can directly show that this superfield expression in fact topological using methods similar to those in \cite{Buchbinder88}. Moreover, $S_{\chi}$ is a combination of the \emph{Euler} and the \emph{Pontryagin} invariant and it has the component expression
\begin{equation}\label{Schicompo}
S_{\chi}=\frac{1}{16}\int d^4x\,e\left(W^{mnpq}W_{mnpq}-2R^{mn}R_{mn}+\frac{1}{6}\mathcal{R}^2\right)+\cdots,
\end{equation}
where $W^{mnpq}$ is the Weyl tensor, $R^{mn}$ is the Ricci tensor and $\mathcal{R}$ is the Ricci scalar. This specific combination appears in the \emph{super Gauss-Bonnet} theorem discussed in for example \cite{Townsend79}.

For the Nielsen-Kallosh ghost, things are slightly different. Its action is given by
\begin{equation*}
S_{\textrm{NK}}=\tr\int d^8z \, EX^{-2}(S+\bar{S})b\bar{b},
\end{equation*}
here the factor $S+\bar{S}$ is absent for the case of a trivial gauge kinetic function. To consider its effect, just as the scenario without a dilaton, we rewrite the action in the form $b\exp(-2V')\bar{b}$ for some $V'$, resembling a super Yang-Mills coupling action. We can absorb this factor by introducing an artificial $U(1)$-factor to our original SYM. This extra $U(1)$ sector has its "gaugino" field given by
\begin{equation}\label{DilU1gaugino}
\mathcal{W}^{\alpha}_{U(1)}=\frac{1}{8}\bar\nabla^2e^{2V'}\nabla^{\alpha}e^{-2V'}=\Delta^\alpha-\frac{2}{3}X^\alpha,
\end{equation}
where 
\begin{equation}\label{Deltadef}
\Delta^\alpha=\frac{1}{8}\bar\nabla^2\nabla^\alpha\log(S+\bar{S})=\frac{1}{8}\bar\nabla^2\left(\frac{\nabla^{\alpha}S}{S+\bar{S}}\right),
\end{equation}
and $X^\alpha$ is the one introduced in \eqref{DilYMdefrecall1} before.

Hence the divergence due to the Nielsen-Kallosh ghost is like that of a free ghost field, which has an extra factor $(-1)$ from its statistics, but with the replacement $\mathcal{W}^{\alpha}_{\textrm{YM}}\rightarrow\mathcal{W}^{\alpha}_{\textrm{YM}}+\mathcal{W}^{\alpha}_{U(1)}$. Using the result of \cite{Butter09}, we have
\begin{equation}\label{DilNKloopdiv}
\begin{aligned}
\Gamma^b_{(1)\textrm{log}}=&\,\frac{\log{\Lambda^2}}{96\pi^2}S_{\chi}\\
&-\frac{\log{\Lambda^2}}{64\pi^2}\left(\left[\left(\mathcal{W}^{\alpha}_{\textrm{YM}}+\Delta^\alpha-\frac{2}{3}X^{\alpha}\right)^2+\frac{2}{3}W^{\alpha\beta\gamma}W_{\gamma\beta\alpha}\right]_F+\textrm{h.c.}\right).
\end{aligned}
\end{equation}

\subsection{Total Logarithmic Divergence}\label{DilSYM4.3}
Now the vector multiplet will have the logarithmic divergence contribution given by
\begin{equation}\label{DilVloopdiv}
\Gamma^V_{(1)\textrm{log}}=\frac{\log{\Lambda^2}}{64\pi^2}\int d^8z\,E[b_2],
\end{equation}
Combining all the results, and taking the trace over the Yang-Mills gauge group with $N_G=\tr 1$ being its rank, the total one-loop logarithmic divergence is given by
\begin{equation}\label{fullDilSYM1looplogdiv}
\begin{aligned}
\Gamma^{(1)}_{\textrm{log}}=&-\frac{3N_G\log{\Lambda^2}}{32\pi^2}[G^aG_a+2R\bar{R}]_D
\\
&-\frac{\log{\Lambda^2}}{64\pi^2}\left(\left[3\tr\mathcal{W}^{\alpha}_{\textrm{YM}}\mathcal{W}_{\textrm{YM},\alpha}+\frac{N_G}{2}X^\alpha X_\alpha+N_GW^{\alpha\beta\gamma}W_{\gamma\beta\alpha}\right]_F+\textrm{h.c.}\right)\\
&+\frac{N_G\log{\Lambda^2}}{32\pi^2}\left[\frac{1}{(S+\bar{S})^2}\left[\nabla^{\alpha}SG_{\alpha\dot{\alpha}}\nabla^{\dot{\alpha}}\bar{S}+\frac{\bar\Sigma\Sigma}{16}+\frac{(\nabla_{\dot{\alpha}}\bar{S})^2(\nabla^{\alpha}S)^2}{8(S+\bar{S})^2}\right]\right]_D\\
&-\frac{\log{\Lambda^2}}{64\pi^2}\left(\left[2\tr\mathcal{W}^{\alpha}_{\textrm{YM}}\Delta_\alpha+N_G\Delta^\alpha\left(\Delta_\alpha-\frac{4}{3}X_\alpha\right)\right]_F+\textrm{h.c.}\right).
\end{aligned}
\end{equation}
Here $\Sigma$ and its conjugate are defined in \eqref{Sigmadef} and the expression for $\Delta^\alpha$ is found in \eqref{Deltadef}. The first two lines are the same as the divergence with a constant coupling strength \cite{Leung19}, and the third and fourth lines are the corrections from introducing the dilaton coupling. It is easy to verify that \eqref{fullDilSYM1looplogdiv} is consistent with the analogous result for the abelian vector multiplet in minimal supergravity \cite{Buchbinder86}. To the best of our knowledge, the result presented here is the first superfield calculation of the one-loop divergence with a non-trivial gauge kinetic function. It will be interesting to compare this with similar results in the literature but with the component approach, for example in \cite{Gaillard97}.

\section{Inclusion of Three Spinor Derivative Terms}\label{DilSYM5}
We have shown that the heat kernel coefficients of a general second order operator can be obtained using a Fourier integration method. In fact, we can go further and apply the same method on an operator with third order derivative terms, but with a restriction: the terms with three derivatives must be constructed only from the spinor derivatives $\nabla^{\alpha}$ and $\nabla_{\dot{\alpha}}$, but not the bosonic ones $\nabla_a$. We shall see how we can incorporate such terms when calculating the heat kernel coefficients.

Let us call the additional third order part of $\mathcal{O}$
\begin{equation}\label{O3rdorderpart}
\mathcal{O}\ni \mathcal{O}^{(3)}=\psi W^{ABC}\nabla_C\nabla_B\nabla_A.
\end{equation}
Here $A$, $B$, $C$ are tensor indices with \emph{only} the spinor part: $A, B, C=\alpha, \dot{\alpha}$, and we have factored out $\psi$ from the coefficients $W^{ABC}$ for simplicity. Now we use the equation \eqref{Fouriershk2}:
\begin{equation*}
\begin{aligned}
K&=\int \frac{d^4k}{(2\pi)^4}e^\phi\exp\left(\sum_{m=0}^{\infty}\frac{(-1)^m}{m!}(\mathcal{L}_\phi)^m(i\tau\mathcal{O})\right)E^{-1}y^\mu y_{\mu}y_{\dot{\mu}}y^{\dot{\mu}}\\
&=\int \frac{d^4k}{(2\pi)^4}e^\phi\exp\left(\sum_{m=0}^{3}\frac{(-1)^m}{m!}(\mathcal{L}_\phi)^m(i\tau\mathcal{O})\right)E^{-1}(y^\mu)^2(y_{\dot{\mu}})^2.
\end{aligned}
\end{equation*}
We now have a term with three commutators, as the operator is of third order. If we rescale $k$ by $k_a\rightarrow k_a\tau^{-1/2}$, the heat kernel coefficients will be given by
\begin{equation}\label{Fourierbn3rd}
\begin{split}
\left[b_n\right]=\frac{n!}{i^{n-1}}\left.\int \frac{d^4k}{\pi^2}\exp\left[A+B\right]E^{-1}(y^\mu)^2(y_{\dot{\mu}})^2\right|_{n, y^M\rightarrow 0},\\
A=i\frac{\mathcal{L}_\phi{}^2}{2}\mathcal{O}, \quad B=i\tau\mathcal{O}-i\tau^{1/2}\mathcal{L}_\phi\mathcal{O}-i\tau^{-1/2}\frac{\mathcal{L}_\phi{}^3\mathcal{O}}{6}.
\end{split}
\end{equation}
Note that the constraints imposed on the third order part of $\mathcal{O}$ imply that in the coincidence limit, $A\rightarrow -ik^2$ and $\mathcal{L}_\phi{}^3\mathcal{O}\rightarrow 0$. To obtain $[b_n]$, one just has to expand the exponential using the formula \eqref{eDysonidentity} and isolate the term proportional to $\tau^n$.

We notice that there is an extra term proportional to $\tau^{-1/2}$, which requires special attention. Without this term in $B$, it is clear that each $[b_n]$, corresponds to $\tau^n$, comes from finitely many number of contributions. This is because each copy of $B$ increases the power of $\tau$ by at least a half, so only terms in the Dyson expansion with less than or equal to $2n$ factors of $B$ will contribute to $[b_n]$. This might not be the case for a third order operator, as $B$ might also decrease the power of $\tau$. Thus we potentially have to deal with an infinite number of terms that will contribute to a particular coefficient $[b_n]$, however we shall argue that this is not the case if we only have spinor derivatives in the third order part of $\mathcal{O}$.

Let us look at the potentially dangerous object $\mathcal{L}_\phi{}^3\mathcal{O}$ in detail. Substituting the expression in \eqref{O3rdorderpart}, we have
\begin{equation}\label{L3Oexpression}
\frac{\mathcal{L}_\phi{}^3\mathcal{O}}{6}=-\psi W^{ABC}\nabla_C\phi\nabla_B\phi\nabla_A\phi,
\end{equation}
which certainly vanishes in the coincidence limit, as each spinor derivative of $\phi$ does. However, when there are extra derivatives acting on $\nabla_A\phi$, the coincidence limit may not vanish. For instance, we have
\begin{equation*}
[\nabla_{\alpha}\nabla_{\dot{\alpha}}\phi]=[-i(\sigma^a)_{\alpha\dot{\alpha}}\nabla_a\phi]=(\sigma^a)_{\alpha\dot{\alpha}}k_a,
\end{equation*}
by using the relation $\{\nabla_{\alpha},\nabla_{\dot{\alpha}}\}=-2i(\sigma^a)_{\alpha\dot{\alpha}}\nabla_a$. As a result, it is possible to have a non-zero coincidence limit for $\mathcal{L}_\phi{}^3\mathcal{O}$ if we have the conjugate derivatives acting on each of the $\nabla\phi$. In other words, in order to have a non-vanishing limit, we need at least three spinor derivatives acting on $\mathcal{L}_\phi{}^3\mathcal{O}$.

In calculating various heat kernel coefficients, we will encounter contributions which contain the functions $f_k[B_1\otimes\cdots\otimes B_k]$, where $B_i$ is either $\mathcal{O}$, $\mathcal{L}_\phi\mathcal{O}$ or $\mathcal{L}_\phi{}^3\mathcal{O}/6$. Suppose we choose one of them to be $\mathcal{L}_\phi{}^3\mathcal{O}/6$. Note that one copy of $\mathcal{O}$ contains terms with exactly three derivatives that can act on $\mathcal{L}_\phi{}^3\mathcal{O}$ to obtain a non-zero limit. When doing so, a copy of $\mathcal{O}$ raises the power of $\tau$ by $\tau^1$ and a copy of $\mathcal{L}_\phi{}^3\mathcal{O}$ lowers the power by $\tau^{-1/2}$, thus we have a net increase in the power of $\tau$ by $\tau^{1/2}$. As for $\mathcal{L}_\phi\mathcal{O}$, it contributes to a power of $\tau^{1/2}$ but contains less than three derivatives. Thus pairing it with $\mathcal{L}_\phi{}^3\mathcal{O}$ will still give a vanishing coincidence limit. 

In conclusion, if we have a factor of $\mathcal{L}_\phi{}^3\mathcal{O}$, there is no way to generate a non-zero result unless it pairs with something that results in a net gain in the power of $\tau$; in fact the power count is raised by at least $\tau^{1/2}$. This implies that for a particular coefficient $[b_n]$, finitely many copies of $\mathcal{L}_\phi{}^3\mathcal{O}$ can be introduced to $f_k[B_1\otimes\cdots\otimes B_k]$ such that it corresponds to $\tau^n$ and has a non-vanishing coincidence limit. Therefore there are only finitely many terms that can contribute to $[b_n]$, which is what we want to prove.

Notice that such an argument will break down if $\mathcal{O}$ contains four or more derivatives, as we will have an extra term proportional to $\tau^{-1}\mathcal{L}_\phi{}^4\mathcal{O}$ and the simple power counting above will not work. Indeed, from the covariant derivative algebra $\{\nabla_{\alpha},\nabla_{\dot{\alpha}}\}=-2i(\sigma^a)_{\alpha\dot{\alpha}}\nabla_a$, the d'Alembertian $\Box$, which provides the kinetic term to the quantum fields and induces the spacetime propagation, is somewhat equivalent to four spinor derivatives. Hence a term with three spinor derivatives will be "less divergent" than the kinetic term, and thus can be treated as a proper perturbation to the free d'Alembertian action. It is no wonder that including terms with three spinor derivatives will provide no trouble but only minor modifications to the calculation of heat kernel coefficients. However, having more than three spinor derivatives will need a different treatment and will not be discussed here.

Let us see briefly how the inclusion of triple spinor derivative terms will affect the calculation of the first three heat kernel coefficients. We always have $[b_0]=0$ from supersymmetry. For $[b_1]$, similar to the previous case we have terms that depend on $f_1[\mathcal{O}]$ or $f_2[\mathcal{L}_\phi\mathcal{O}\otimes\mathcal{L}_\phi\mathcal{O}]$. For the former one, recall that we need at least four spinor derivatives to annihilate the factor $(y^\mu)^2(y_{\dot{\mu}})^2$ in order to have a non-zero coincidence limit, we see that $f_1[\mathcal{O}]$ cannot contribute as it is of third order. Now for $f_2[\mathcal{L}_\phi\mathcal{O}\otimes\mathcal{L}_\phi\mathcal{O}]$, notice that $\mathcal{L}_\phi\mathcal{O}$ is of the form
\begin{equation}\label{3rdorderLO}
\mathcal{L}_\phi\mathcal{O}=\psi \tilde{W}^{ABC}(\nabla_C\phi)\nabla_B\nabla_A+\textrm{lower order terms}.
\end{equation}
We immediately see that lower order terms cannot contribute as there are not enough derivatives, and the only four spinor derivative terms in $f_2[\mathcal{L}_\phi\mathcal{O}\otimes\mathcal{L}_\phi\mathcal{O}]$ will depend on $\nabla_C\phi$, and thus the coincidence limit vanishes. In short, $f_2[\mathcal{L}_\phi\mathcal{O}\otimes\mathcal{L}_\phi\mathcal{O}]$ cannot generate a non-zero $[b_1]$.

Next we may have some new contributions due to the existence of the new term $\mathcal{L}_\phi{}^3\mathcal{O}/6$. For example, $f_3[\mathcal{O}\otimes\mathcal{L}_\phi{}^3\mathcal{O}/6\otimes]\mathcal{L}_\phi\mathcal{O}$ has the power count being $\tau^1$, which may contribute to $[b_1]$. However, counting the number of derivatives shows that there cannot be any non-zero result. It is similar for other potential contributions involving $\mathcal{L}_\phi{}^3\mathcal{O}/6$, so we still have $[b_1]=0$.

The next coefficient $[b_2]$ will be more interesting, and let us look at some of the old terms. We first have one that depends on $f_2[\mathcal{O}\otimes\mathcal{O}]$, which is now of sixth differential order.
As we only need four spinor derivatives for a non-zero coincidence limit, various extra features arise. First, the linear part of $\mathcal{O}$ will contribute, as it can pair with the cubic part to get four derivatives. Previously only the quadratic part of $\mathcal{O}$ matters, and now we have also the first order part to take into account. However the non-derivative part will still be irrelevant, in particular the mass term will have no effect. 

Second, as there can be a six derivative term, there are two derivatives that can act on $E^{-1}$ when taking the coincidence limit; thus we will need its normal coordinate expansion up to second order. In the old case, we will not need such a expansion as we have four derivatives at maximum, and the zeroth order expansion of $E^{-1}$ is just one. Also, we may have some derivatives of the first $\mathcal{O}$ acts on the second $\mathcal{O}$, so the final result may depend on derivatives of the coefficients of $\mathcal{O}$, whereas previously $[b_2]$ is only an algebraic expression with no derivatives, as in \eqref{generalO2b2}.

Third, in $f_2[\mathcal{O}\otimes\mathcal{O}]$, we will encounter the term $\mathcal{L}_A{}^m\mathcal{O}$, which appears when commutating the exponentials involving $A$ past the operator $\mathcal{O}$. For the previous setup without the third order term, we are forced to choose $m=0$ as otherwise there will not be enough derivatives for a non-zero result. But now we have two spinor derivatives in surplus so we can take $m$ to be at most two. As a result, upon the $k$ integration we will have a term proportional to $\mathcal{L}_\psi{}^2\mathcal{O}$, thus the second derivative of $\psi$ will appear in $[b_2]$, which of course does not happen for the old case.

Now let us analyze the term $f_3[\mathcal{O}\otimes\mathcal{L}_{\phi}\mathcal{O}\otimes\mathcal{L}_{\phi}\mathcal{O}]$. From the form of $\mathcal{L}_{\phi}\mathcal{O}$ as in \eqref{3rdorderLO}, we see that its quadratic part contains the expression $\nabla_C\phi\nabla_B\nabla_A$. This term is roughly equivalent to one derivative, as we need an extra derivative to act on $\nabla_C\phi$ for a non-zero coincidence limit, thus we have a net gain of one derivative as a result. Therefore $\mathcal{L}_{\phi}\mathcal{O}$ is similar to a linear operator, and so $f_3[\mathcal{O}\otimes\mathcal{L}_{\phi}\mathcal{O}\otimes\mathcal{L}_{\phi}\mathcal{O}]$ is like a fifth order operator. For a term with five derivatives, we will need the first order normal coordinate expansion of $E^{-1}$, which will be trivial if the trace of the torsion vanishes, $T_{AB}{}^B=0$, as in commonly seen theories. We will also encounter the first derivative of $\psi$ in the final result after the integration over $k$. There are two more terms involving $f_3$, and they will be similar. Finally the term with $f_4[\mathcal{L}_{\phi}\mathcal{O}^{\otimes{4}}]$ will have no significant difference from the old case.

We might also have new contributions that include $\mathcal{L}_\phi{}^3\mathcal{O}/6$. A simple inspection shows that there are new terms that depends on $f_4[\mathcal{O}\otimes\mathcal{O}\otimes\mathcal{L}_\phi\mathcal{O}\otimes\mathcal{L}_\phi{}^3\mathcal{O}/6]$ and similar terms with the operators permuted. Such an operator is roughly a fourth order one, however we will need the precise form of $\mathcal{O}$ to see how this contributes to $[b_2]$. For terms with two or more copies of $\mathcal{L}_\phi{}^3\mathcal{O}/6$, counting the number of derivatives shows that they cannot contribute to $[b_2]$, thus the one shown above is the only contribution that includes $\mathcal{L}_\phi{}^3\mathcal{O}/6$.

This concludes the discussion of $[b_2]$, and we can similarly analyze the higher order heat kernel coefficients as above. In general for $[b_n]$, we will come across operators of at most $3n$ differential order, up from $2n$ as in the old case. This implies that we will in general need the $(3n-4)$-th order normal coordinate expansion, and the final answer will contains $(3n-4)$-th derivatives of the coefficients of $\mathcal{O}$. There will be terms that depend on $\mathcal{L}_\phi{}^3\mathcal{O}/6$. Simple power counting shows that there can be at most $2n-3$ copies of $\mathcal{L}_\phi{}^3\mathcal{O}/6$ introduced. In fact, including the extra third order spinor derivative term of $\mathcal{O}$ merely increases the amount of algebra involved to calculate $[b_n]$. The previous method for second order operators applies equally well here for these special third order operators, without much difficulty introduced.

\section{Conclusion}
We have developed a Fourier integral technique for calculating the heat kernel coefficients, applicable for any second order operators and some special third order ones. Using the general result, we have derived the one-loop divergence of the dilaton-coupled super Yang-Mills theory. The result presented is quite general, we may readily apply it for different theories with different field contents. For instance, while the linear multiplet at one-loop level in supergravity is discussed in the literature \cite{Grisaru84,Buchbinder88}, the \emph{modified} linear multiplet was not considered. This modified version has certain phenomenological interest, as it enables a non-homomorphic SYM gauge coupling \cite{BGG91,BBG00} which typically arises from string induced models. Another promising candidate to study is the quanta of the gravitational multiplet, which is a gauge vector multiplet with an extra bosonic index $V^a$ \cite{Siegel79,Grisaru81,Grisaru84,Butter10b}. Studying this will allow us to examine quantized supergravity at one-loop level.

Instead of staying within $N=1$ superspace in four dimension, we might also consider different theories with different superspaces. For example, $N=2$ supergravity is an active area of study. It is hoped that one can generalize the technique presented here to the case of $N=2$ superspace. We might even go beyond and consider supersymmetric theories in different dimension, for example those in string theory. Such generalization will be a subject of interest.

Finally, we have restricted ourselves to second order operators. One may ask how heat kernel coefficients change if general higher order derivative terms are introduced. This can be analyzed using perturbation theory for heat kernel, and will be considered in future work.

\section*{Acknowledgments}
The author would like to thank Mary K. Gaillard for helpful discussions and comments. This work was supported in part by the Director, Office of Science, Office of High Energy and Nuclear Physics, Division of High Energy Physics, of the US Department of Energy under Contract DE-AC02-05CH11231 and in part by the National Science Foundation under grant PHY-1316783.

\newpage
\appendix
\appendixpage
\section{Conformal Superspace and Quantization of SYM Theory in Conformal SUGRA}
In this appendix, we shall review the superspace formalism of conformal supergravity, developed by Butter \cite{Butter10}, and the quantization of super Yang-Mills theory with constant coupling in conformal supergravity, presented in \cite{Leung19}.

The conformal superspace is an $N=1$ superspace with the \emph{superconformal algebra}, which is generated by the operators $\{P_A,M_{ab},D,A,K_A\}$.\footnote{Here the subscript $A$ runs over the indices $\{a,\alpha,\dot{\alpha}\}$.} Here $D$ is the dilatation, $A$ is the chiral rotation, and $K_A$ are the special conformal transformations. The commutation relation of these generators can be found in the original reference \cite{Butter10}. We may construct the covariant derivative by introducing a connection for every generators except $P_A$:
\begin{equation}\label{covderirev}
\nabla_M=\partial_M-\frac{1}{2}\phi_M{}^{ba}M_{ab}-B_MD-A_MA-f_M{}^AK_A.
\end{equation}
We also introduce the supervielbein $E_M{}^A$, which allows us to interchange between an Einstein index and a Lorentz index. Then, the action of $P_A$ on a scalar $\Phi$ is the same as the covariant derivative $\nabla_A$:
\begin{equation}\label{P=covdrev}
P_A\Phi=\nabla_A\Phi=E_A{}^M\nabla_M\Phi.
\end{equation}

The graded commutator of $P_A$ determines the curvature:
\begin{equation}\label{defromedPrev}
\left[P_A,P_B\right]=-T_{AB}{}^CP_C-\frac{1}{2}R_{AB}{}^{dc}M_{cd}-H_{AB}D-F_{AB}A-R(K)_{AB}{}^CK_C.
\end{equation}
Here the curvature tensors like $R_{AB}$, $H_{AB}$ can be calculated from the connection fields. By imposing suitable constraints \cite{Butter10} on the torsion tensor and the curvature tensors, one can actually solve the Bianchi identity and obtain a consistent solution, similar to the case of ordinary supergravity. In particular, the covariant derivative algebra is quite simple:
\begin{equation}\label{covalgebrarev}
\begin{aligned}
&\{\nabla_{\alpha},\nabla_{\beta}\}=\{\bar{\nabla}_{\dot{\alpha}},\bar{\nabla}_{\dot{\beta}}\}=0,\\
&\{\nabla_{\alpha},\bar{\nabla}_{\dot{\beta}}\}=-2i\nabla_{\alpha\dot{\beta}}, \\
&[\nabla_{\alpha},\nabla_{{\beta}\dot{\beta}}]=-2i\epsilon_{\alpha\beta}\mathcal{W}_{\dot{\beta}}, \quad [\nabla_{\dot{\alpha}},\nabla_{{\beta}\dot{\beta}}]=-2i\epsilon_{\dot{\alpha}\dot{\beta}}\mathcal{W}_{\beta}.\\
&[\nabla_{\alpha\dot{\alpha}},\nabla_{\beta\dot{\beta}}]=\epsilon_{\dot{\alpha}\dot{\beta}}\{\nabla_\alpha,\mathcal{W}_{\beta}\}+\epsilon_{\alpha\beta}\{\bar{\nabla}_{\dot{\alpha}},\mathcal{W}_{\dot{\beta}}\}.
\end{aligned}
\end{equation}
Here $\mathcal{W}_{\alpha}$ and its conjugate are the "gaugino" superfield, which is given by
\begin{equation}\label{Wcomponentrev}
\mathcal{W}_{\alpha}=\frac{1}{2}\mathcal{W}(M)_{\alpha}{}^{cb}M_{bc}+\mathcal{W}(K)_{\alpha}{}^{A}K_A,
\end{equation}
where $\mathcal{W}(M)_{\alpha}{}^{cb}$ and $\mathcal{W}(K)_{\alpha}{}^{A}$ can be expressed in terms of a symmetric super-Weyl tensor $W^{\alpha\beta\gamma}$, the details can be found in \cite{Butter10}. The gaugino superfield also satisfies the chirality condition and the Bianchi identity:
\begin{equation}\label{guaginoconrev}
\begin{aligned}
&\{\nabla_\alpha,\mathcal{W}_{\dot{\beta}}\}=\{\bar{\nabla}_{\dot{\alpha}},\mathcal{W}_{\beta}\}=0\\
&\{\nabla^\alpha,\mathcal{W}_{\alpha}\}=\{\bar{\nabla}_{\dot{\beta}},\mathcal{W}^{\dot{\beta}}\}.
\end{aligned}
\end{equation}

Matters in a conformal supergravity theory are described by \emph{primary superfields}. A primary superfield $\phi$ must satisfy $K_A\phi=0$, and has a conformal weight $(\Delta,w)$ if $D\phi=\Delta\phi$ and $A\phi=iw\phi$. \footnote{Note that there may be extra restrictions depending on the type of field being considered, for instance a primary \emph{chiral} field must have $3\Delta=2w$.} These primary fields can be used to construct $D$-term and $F$-term actions. A $D$-term and an $F$-term must have conformal weights $(2,0)$ and $(3,2)$ respectively. They can be converted to each other by using the \emph{chiral projector} $\mathcal{P}=-\bar\nabla^2/4$.

It is important to note that integration by parts in conformal superspace is actually non-trivial, as the property of being conformal primary is \emph{not} preserved when taking a covariant derivative. As a result an extra correction term is introduced, the modified integration by parts formula is given by
\begin{equation}\label{bypartsrev}
E\nabla_Av^A=\nabla_M(EE_A{}^Mv^A)-Ef_{A}{}^BK_Bv^A,
\end{equation}
where $f_{A}{}^B$ is the connection corresponds to the special conformal transformation. Therefore up to a total derivative, we have 
\begin{equation}\label{bypartsrev2}
\nabla_Av^A=-f_{A}{}^BK_Bv^A.
\end{equation}
More details can be found in the appendix of \cite{Leung19}.

It is easy to introduce Yang-Mills theory, with gauge generators $\{X_{(r)}\}$, to conformal supergravity. One just needs to introduce an extra Yang-Mills contribution to the covariant derivative and the guagino:
\begin{equation}\label{YMcovWrev}
\begin{aligned}
\nabla_A&\rightarrow\nabla_A-\mathcal{A}^{(r)}_AX_{(r)}\\
\mathcal{W}_\alpha&\rightarrow\mathcal{W}_\alpha+\mathcal{W}_{\alpha,\textrm{YM}}=\mathcal{W}_\alpha+\mathcal{W}^{(r)}_{\alpha}X_{(r)}.
\end{aligned}
\end{equation}
Here $\mathcal{A}^{(r)}_A$ is the Yang-Mills gauge connection and the Yang-Mills gaugino $\mathcal{W}_{\alpha,\textrm{YM}}$ can be calculated in terms of it. The Yamg-Mills action is
\begin{equation}\label{conformalYMactionrev}
S_{\textrm{YM}}=\frac{1}{4}\int d^4xd^2\theta \, \mathcal{E} f_{(r)(s)}\mathcal{W}^{(r)\alpha}\mathcal{W}^{(s)}_{\alpha}+\textrm{h.c.},
\end{equation}
with $f_{(r)(s)}$ being the \emph{gauge kinetic function}. In the following we shall consider the simplest case which $f_{(r)(s)}=g^{-2}\delta_{(r)(s)}$.

We shall employ the background field method to quantize the theory, which is similar to the case in flat superspace.\footnote{One may, for instance, consult \cite{Gates83} for details of the flat scenario.} We perform the background-quantum splitting by introducing the Yang-Mills prepotential
\begin{equation}\label{defprepo}
\nabla_{\alpha}=S_Q^{-1}\nabla_{B\alpha}S_Q, \quad \nabla_{\dot{\alpha}}=T_Q^{-1}\nabla_{B\dot{\alpha}}T_Q,
\end{equation}
where $\nabla_{B}$ denotes the background covariant derivative, and $S_Q$, $T_Q$ are the quantum prepotential. We define the vector multiplet $V$, the quanta of the theory, via 
\begin{equation}\label{Vqdef}
U_Q=S_QT_Q^{-1}=\exp(-2iV).
\end{equation}

To fix the gauge freedom, we shall choose the following conditions:
\begin{equation}\label{YMgaugerev}
\bar\nabla^2(XV)-f=0, \quad \nabla^2(XV)-\bar{f}=0.
\end{equation}
Here $X$ is the so-called \emph{compensator}, with conformal weights $(2,0)$, which is introduced in order to make the gauge condition \emph{conformal primary}.
The gauge-fixing action we use will be 
\begin{equation}\label{gfactionrev}
S_{\textrm{g.f.}}=\frac{1}{8g^2}\tr\int d^8z \,EX^{-2}[\bar\nabla^2(XV)\nabla^2(XV)],
\end{equation}
note that we have to include a factor of $X^{-2}$ for a valid $D$-term action.
There are extra ghost fields introduced from the gauge-fixing procedure, one of which is the Feddeev-Popov ghost with the action being
\begin{equation}\label{FPghostactionrev}
S_{\textrm{FP}}=\tr\int d^8z \, EX(c'+\bar{c}')\mathcal{L}_{V/2}[c-\bar{c}+\coth(\mathcal{L}_{V/2})(c+\bar{c})],
\end{equation} 
here $\mathcal{L}_{V/2}f=[V/2,f]$ is the commutator. Another ghost we have is the Nielsen-Kallosh ghost, its action is
\begin{equation}\label{ghostaction}
S_{\textrm{NK}}=\tr\int d^8z \, EX^{-2}b\bar{b}.
\end{equation} 

In order to analyze the theory at one-loop, it is necessary to expand the gauge-fixed action to the second order in $V$. With some calculation, including a careful treatment when performing integration by parts, it can be shown \cite{Leung19} that the result is given by
\begin{equation}\label{simplifyaction2ndbrev}
S_{\textrm{YM}}^{(2)}=\frac{1}{2}\tr\int d^8z\, E\left(\frac{2}{g^2}\right)V\mathcal{O}_VV,
\end{equation}
where
\begin{equation}\label{YMOVrev}
\begin{aligned}
\mathcal{O}_V&=\Box+\frac{1}{2}G^{\alpha\dot{\alpha}}[\nabla_{\alpha},\nabla_{\dot\alpha}]+\left(\frac{X^\alpha}{3}-\nabla^{\alpha}R+\mathcal{W}_{\textrm{YM}}^{\alpha}\right)\nabla_{\alpha}\\
&+\left(\frac{X_{\dot\alpha}}{3}-\nabla_{\dot\alpha}\bar{R}-\mathcal{W}_{\textrm{YM},\dot\alpha}\right)\nabla^{\dot{\alpha}}-\frac{1}{2}\left(\bar\nabla^2\bar{R}+\nabla^2R+16R\bar{R}\right)\\
&+\frac{i}{4}U^\alpha(\nabla^{\dot{\beta}}\nabla_{\alpha\dot{\beta}}+\nabla_{\alpha\dot{\beta}}\nabla^{\dot{\beta}})+\frac{i}{4}U_{\dot{\alpha}}(\nabla^{\beta\dot{\alpha}}\nabla_{\beta}+\nabla_{\beta}\nabla^{\beta\dot{\alpha}})\\
&+\frac{1}{8}\left(U_{\dot{\alpha}}U^{\dot{\alpha}}\nabla^2+U^{\alpha}U_{\alpha}\bar\nabla^2+4U^{\alpha}U^{\dot{\alpha}}[\nabla_{\alpha},\nabla_{\dot\alpha}]\right)\\
&+\frac{1}{4}\left(8RU^\alpha+U_{\dot{\alpha}}U^{\dot{\alpha}}U^\alpha-U_{\dot\alpha}U^{\dot{\alpha}\alpha}\right)\nabla_\alpha\\
&+\frac{1}{4}\left(8\bar{R}U_{\dot{\alpha}}+U^{\alpha}U_{\alpha}U_{\dot{\alpha}}-U^{\alpha}U_{\alpha\dot{\alpha}}\right)\nabla^{\dot{\alpha}}\\
&+U^a\nabla_a+\left(U^{\alpha}\nabla_{\alpha}R+U_{\dot{\alpha}}\nabla^{\dot{\alpha}}\bar{R}-U^{\alpha}U_{\alpha}R-U_{\dot{\alpha}}U^{\dot{\alpha}}\bar{R}\right).
\end{aligned}
\end{equation}
The new fields introduced here are defined by
\begin{equation}\label{YMdefrev}
\begin{aligned}
U^\alpha=\nabla^{\alpha}\log{X}, &\quad U_{\dot{\alpha}}=\nabla_{\dot\alpha}\log{X},\\
R=-\frac{1}{8X}\bar\nabla^2X, &\quad\bar{R}=-\frac{1}{8X}\nabla^2X,\\
X_\alpha=\frac{3}{8}\bar\nabla^2U_{\alpha}, &\quad X^{\dot{\alpha}}=\frac{3}{8}\nabla^2U^{\dot\alpha},\\
G_{\alpha\dot{\alpha}}=-\frac{1}{4}(U_{\alpha\dot{\alpha}}&-U_{\dot{\alpha}\alpha})-\frac{1}{2}U_{\alpha}U_{\dot{\alpha}},\\
U_{\alpha\dot{\alpha}}=\nabla_{\alpha}U_{\dot{\alpha}}, &\quad U_{\dot{\alpha}\alpha}=\nabla_{\dot\alpha}U_{\alpha}.
\end{aligned}
\end{equation}
It is noted that upon the conformal gauge fixing $U^\alpha=U_{\dot{\alpha}}=0$, the fields $R$, $\bar{R}$, $G_{\alpha\dot{\alpha}}$, $X_\alpha$ and $X^{\dot{\alpha}}$ reduce to those with the same name in $U(1)$-supergravity \cite{Butter10b}.

\section{An Example of Calculating Fourier Integration of Operators in Heat Kernel Coefficient Calculations}
In the following, we shall consider a concrete example of how to compute the coincidence limit of certain Fourier integrals that are related to heat kernel coefficients.

As an example, we consider an operator of the form
\begin{equation}\label{ExampleO}
\mathcal{O}=\psi\Box+F\nabla^2+\bar{F}\bar\nabla^2+Q.
\end{equation} 
We then have 
\begin{equation}\label{ExampleLO}
\mathcal{L}_\phi\mathcal{O}=-[\mathcal{O},\phi]=-2\psi\nabla^a\phi\nabla_a-2F\nabla^\alpha\phi\nabla_\alpha-2\bar{F}\nabla_{\dot{\alpha}}\phi\nabla^{\dot{\alpha}}-C,
\end{equation}
where $C=\psi\Box\phi+F\nabla^2\phi+\bar{F}\bar\nabla^2\phi$. Note that $\nabla^A\phi$ has the coincidence limit $[\nabla^A\phi]=ik^a\delta_a{}^A$, and $[C]=0$. Next we have 
\begin{equation}\label{ExampleL2O}
\frac{(\mathcal{L}_\phi)^2}{2}\mathcal{O}=\psi\nabla^a\phi\nabla_a\phi+F\nabla^\alpha\phi\nabla_\alpha\phi+\bar{F}\nabla_{\dot{\alpha}}\phi\nabla^{\dot{\alpha}}\phi,
\end{equation}
its coincidence limit being $[\frac{(\mathcal{L}_\phi)^2}{2}\mathcal{O}]=-\psi k^ak_a$ as expected.

Let us calculate
$\int\frac{d^4k}{\pi^2}f_1[\mathcal{O}]$ and $\int\frac{d^4k}{\pi^2}f_2[\mathcal{L}_\phi\mathcal{O}\otimes\mathcal{L}_\phi\mathcal{O}]$, which appear in the calculation of $[b_1]$. For the first one, we have
\begin{equation}\label{examplef1}
\begin{aligned}
f_1[\mathcal{O}]&=\int_0^1d\alpha_1\,e^{\alpha_1A}\mathcal{O}e^{(1-\alpha_1)A}\\
&=\int_0^1d\alpha_1\,e^{\alpha_1A}e^{(1-\alpha_1)A}e^{\mathcal{L}_{-(1-\alpha_1)A}}\mathcal{O}\\
&=\int_0^1d\alpha_1\,e^{A}\sum_{m=0}^\infty\frac{(1-\alpha_1)^m}{m!}(\mathcal{L}_{-A})^m\mathcal{O}\\
&=e^{A}\sum_{m=0}^2\frac{1}{(m+1)!}(\mathcal{L}_{-A})^m\mathcal{O},
\end{aligned}
\end{equation}
the summation terminates at $m=2$ as $\mathcal{O}$ is of second order.

Now we integrate over $k$, in the coincidence limit:
\begin{equation}\label{examplef1kint}
\begin{aligned}
\left.\int\frac{d^4k}{\pi^2}f_1[\mathcal{O}]\right|&=\int\frac{d^4k}{\pi^2}e^{-i\psi k^2}\sum_{m=0}^2\frac{(ik^2)^m}{(m+1)!}(\mathcal{L}_{\psi})^m\mathcal{O}\\
&=-i\sum_{m=0}^2\int dx\,e^{-\psi x}\frac{x^{m+1}}{(m+1)!}(\mathcal{L}_{\psi})^m\mathcal{O}\\
&=-i\sum_{m=0}^2\psi^{-(m+2)}(\mathcal{L}_{\psi})^m\mathcal{O}.
\end{aligned}
\end{equation}
In the second line, we have used a Wick rotation: $x=ik^2$ and integrated over the 4D-hypersphere. Notice that this expression contains various derivative terms, as $\mathcal{O}$ and $\mathcal{L}_{\psi}\mathcal{O}$ are respectively second and first order differential operators.

For the term $\int\frac{d^4k}{\pi^2}f_2[\mathcal{L}_\phi\mathcal{O}\otimes\mathcal{L}_\phi\mathcal{O}]$, the idea is similar, and the details will be omitted here. We have to move the exponential past $\mathcal{L}_\phi\mathcal{O}$ twice, which results in the factor
$(\mathcal{L}_{-A})^m\mathcal{L}_\phi\mathcal{O}(\mathcal{L}_{-A})^n\mathcal{L}_\phi\mathcal{O}$.  Since $\mathcal{L}_\phi\mathcal{O}$ is of first order, we must have $n\le1$ and $m\le2-n$ for a non-zero result. After performing the $\alpha$ integral and going to the coincidence limit, we have to evaluate the integral of the form:
\begin{equation*}
\int\frac{d^4k}{\pi^2}e^{-i\psi k^2}k^ak^b(ik^2)^{(m+n)}(\mathcal{L}_{\psi})^m\psi\nabla_a(\mathcal{L}_{\psi})^n\psi\nabla_b.
\end{equation*}
We can by replace $k^ak^b$ by $\eta^{ab}k^2/4$ using symmetry arguments. Then the $k$-integral can be calculated similarly to the previous case. The final result is
\begin{equation}
\int\frac{d^4k}{\pi^2}f_2[\mathcal{L}_\phi\mathcal{O}\otimes\mathcal{L}_\phi\mathcal{O}]=\sum_{n=0}^{1}\sum_{m=0}^{2-n}C_{m,n}\psi^{-(m+n+2)}(\mathcal{L}_{\psi})^m\nabla^a(\mathcal{L}_{\psi})^n\psi\nabla_a,
\end{equation}
with $C_{m,n}$ some constant that can be easily determined case by case, as $m$ and $n$ are small numbers here.

Note that instead of a specific $\mathcal{O}$ as in \eqref{ExampleO}, the treatment for a more general second order operator is similar. Hence one can, with the recipe outlined here, actually find the closed form expression for this class of Fourier integrals.

\end{document}